%% file: main_jmp.tex
\documentclass[12pt,a4papper,fleqn,titlepage,oneside]{article}
\usepackage[english]{babel}
\usepackage[dvips]{graphicx}
\usepackage{hhline,xspace,exscale}
\usepackage{amsfonts,amssymb,amsgen,amsbsy}
\usepackage{latexsym}
\addtocounter{page}{-1}
\usepackage[centertags]{amsmath}
\numberwithin{equation}{section}
\title{Linear stability analysis of hedgehogs in the
Skyrme model on the three-sphere. \\Critical phenomena and
spontaneously broken reflection symmetry.}
\author{{\L}ukasz Bratek\\ {\scriptsize
Centre for Particle Theory,
Department of Mathematical Sciences,}\\
{\scriptsize University of Durham, Durham, DH1 3LE, UK}\\ {\scriptsize {\it e-mail:} lukasz.bratek@durham.ac.uk} \\{\scriptsize \& Jagellonian University, Reymonta 4, Cracow, Poland}}
\date{may 2004}
\begin{document}
\input{jmp_dfn.tx}
\maketitle
\input{jmp_abs.tx}
\input{jmp_skr.tx}
\input{jmp_stg.tx}
\input{jmp_sts.tx}
\section{Conclusions}
\input{jmp_con.tx}
%
\newpage
\appendix
\input{jmp_app.tx}
\newpage

\input{jmp_bib.tx}
\newpage
\tableofcontents 
\end{document}

%% file: jmp_abs.tx
\begin{abstract}
This paper is devoted to the linear stability analysis of the
whole spectrum of static hedgehog solutions of the Skyrme model on
the three-sphere of  radius $L$. These solutions are described by
profiles $F(\psi)$ that satisfy the equation

\begin{scriptsize}
\begin{eqnarray*}
\left(L+\frac{2}{L}\frac{\sin ^2{F(\psi)}}{\sin ^2{\psi }}\right)\sin^2{\psi }F''(\psi)+
\left(L+\frac{1}{L}\frac{\sin{2F(\psi)}}{\sin{2\psi}}F'(\psi)\right)
\sin{2\psi }F'(\psi)\\-\left(L+\frac{1}{L} \frac{\sin
^2{F(\psi)}}{\sin ^2{\psi }}\right)\sin{2F(\psi)}=0,
\end{eqnarray*}
\end{scriptsize}

\noindent
where
$\psi\in[0,\pi]$, $F(0)=0$, $ F(\pi)=\q\pi$ and $\q$ is an integer
which has the interpretation of  topological charge. This work is
the continuation of the paper \cite{my} where these solutions were
classified. It turns out that only solutions that in the limit
$L\to \infty$ tend to skyrmions (localized at the poles) are
linearly stable. The other solutions are unstable and, for a given
solution, the number of instabilities, for $L$ sufficiently large,
is equal to the index of a harmonic map to which this solution
tends pointwise in the limit $L\to\infty$. Such  solutions, which
in addition have a definite parity, undergo a transition by $+1$
in the number of instabilities as $L$ grows. Due to the
instability, new solutions, with spontaneously broken reflection
symmetry, are born by bifurcations. In the case of the
$1$-skyrmion this critical phenomenon can be fully described
analytically. This allows of prediction the unique series
expansions for the profile of the $1$-skyrmion and for its energy,
though their expansion coefficients are not given in a general
form, and read respectively

\begin{scriptsize}
\begin{eqnarray*} F(\psi)&=&\psi+x\sin{\psi}+x^2\frac{3}{20}\sin{2\psi}
-x^3\left(\frac{29369}{316800}\sin{\psi}-\frac{11}{480}\sin{3\psi}\right)
+\olarge{(x^4)}
\\
 \frac{E[F]}{12\pi^2}&=&\frac{3\sqrt{2}}{4}\left(1+\frac{11}{180}x^2
-\frac{209}{10800}x^4+\frac{5209}{864000}x^6
+\olarge{(x^6)}\right), \qquad
x=\pm\sqrt{\frac{60}{11}\left(\frac{L}{\sqrt{2}}-1\right)}.
\end{eqnarray*}
\end{scriptsize}

\noindent
The above expansions are valid in some right neighbourhood of the
critical radius $L=\sqrt{2}$ at which the $1$-skyrmion (with
broken reflection symmetry) is born by a 'costing no energy'
excitation of the marginal stability mode of parity $+1$ on the
identity solution, which has parity $-1$. To the author's best
knowledge the series were not given in literature
so far. A similar mechanism of
spontaneous breaking of parity is also observed when other
solutions appear by bifurcations from symmetric solutions.
\end{abstract}

%% file: jmp_skr.tx
\section{The Skyrme model on $\s{3}$}
The Skyrme model is defined by the action which written in
generally covariant form reads
\begin{equation}\mathcal{S}[U]=
\int \sqrt{-g}\ud^4x\left(\frac{1}{4}\f^2\Tr\left(K _\mu
K^\mu\right) +\frac{1}{32\e^2}\Tr\left([K_\mu ,K_\nu][K^\mu
,K^\nu]\right)\right) \label{eq:lagdens},\end{equation} where $U$
is the basic chiral $SU(2)$-valued scalar field and $K_\mu=\imath
U^+\partial _\mu U$ is a Lie algebra valued four-current. We use
the convention $(+,-,-,-)$ for the signature of the canonical
quadratic form associated with a metric tensor. The first
integrand is the $\sigma$-term, the second one was introduced by
Skyrme in \cite{skyrme3} to ensure the existence of solitons, but
its shape may be also determined by using geometrical arguments
\cite{manton87},\cite{lukierski}. Both these terms have mutually
inverse scale dependence, thus compete against each other. Unlike
in the Skyrme model on flat space, in addition to the natural size
of a soliton $(\e\f)^{-1}$, we have another natural size at our
disposal which is introduced by the radius $R$ of the base
three-sphere. Their quotient $L=\e\f R$ (which is a dimensionless
number as the $\pi$) must have nontrivial consequences to our
model since the parameter $L$, as a remnant of the mentioned
competition, enters the equations of motion.

In \cite{my} we found and classified all static, finite energy,
spherically symmetric and equivariant mappings $F$ from the base
space -- the three-sphere of radius $R$, to the target space --
the group $SU(2)$. These mappings are regular solutions of the
equation
 \begin{eqnarray} \left(L+\frac{2}{L}\frac{\sin ^2{F(\psi)}}{\sin ^2{\psi
}}\right)\sin ^2{\psi }F''(\psi)+\nonumber\\
\left(L+\frac{1}{L}\frac{\sin{2F(\psi)}}{\sin{2\psi}}F'(\psi)\right)\sin{2\psi
}F'(\psi)-\nonumber\\\left(L+\frac{1}{L} \frac{\sin ^2{F(\psi)}}{\sin ^2{\psi
}}\right)\sin{2F(\psi)}=0,\label{eq:main}
\end{eqnarray}
 which is singular on boundaries.
We utilized the canonical correspondence between points
$(\Psi,\Theta,\Phi)$ on $\s{3}$ and elements of $SU(2)$ (the last
is metrically equivalent to $\s{3}$), that is if $U\in SU(2)$ then
$$U(\Psi,\Theta,\Phi)=\left(\begin{array}{cc}
\cos{\Psi}+\imath\sin{\Psi}\cos{\Theta}&\imath\sin{\Psi}\sin{\Theta}e^{-\imath \Phi}\\
\imath \sin{\Psi}\sin{\Theta} e^{\imath
\Phi}&\cos{\Psi}-\imath\sin{\Psi}\cos{\Theta}\end{array}\right).$$
The base three sphere is endowed with the standard spherical
coordinates $(\psi, \vartheta,\varphi)$ in which the line element
reads
$$\ud{s}^2=R^2\br{\ud{\psi}^2+ \sin^2{\psi}\br{\ud{
\vartheta}^2+\sin^2{\vartheta}\ud{\varphi}^2}},\quad
\{\psi,\vartheta,\varphi\}\in [0,\pi]\times [0,\pi]\times
[0,2\pi].$$ The equivariance of the mapping $\s{3}\to\s{3}$ means
that \begin{equation}F:\quad \s{3}\ni
(\psi,\vartheta,\varphi)\rightarrow
(\Psi=F(\psi),\Theta=\vartheta,\Phi=\varphi)\in \s{3},
\label{eq:map}
\end{equation} thus is related to the Skyrme's ansatz,
which is popularly called  the 'hedgehog' ansatz.

The number of solutions of equation (\ref{eq:main}) increases as L
grows. In the limit $L\to\infty$, a given solution  tends to a
configuration which is composed of two multi-skyrmions $\sk{n}$ and
$\sk{m}$, respectively, localized at the north and at the south
pole of the base three-sphere, and a 'harmonic map' $\ha{p}$ in
between. These solutions are denoted symbolically  by
$\sk{n}\ha{p}\sk{m}$. This means that each solution tends
pointwise to a harmonic map $\has{p}$ of \bizon \cite{biz} in some
interval $\varepsilon<\psi<\pi-\varepsilon$, and $\varepsilon\to0$
as $L\to\infty$. If, in the vicinity of the north pole, one
introduces the radial variable $r$ defined by $1\gg\psi\propto
rL^{-1}$, then $F(r)$ approximately satisfies the equation  of the
Skyrme model on flat space \cite{mat}. In the vicinity of the
south pole one can proceed in an analogous way.

In what follows, for convenience, we will often refer to $F'(0)$
and $F'(\pi)$ as shooting parameters and also to be in connection
with \cite{my} where this nomenclature was introduced due to the
specific method used for finding solutions.

The solutions were examined to see which static configurations of
the field $F$ are possible at a given size of the base
three-sphere. Energies of these configurations are given by the
functional
\begin{eqnarray}&& \mathcal{U}[F]=\int \limits_{0}^\pi w(\psi)U(\psi,
F(\psi),F'(\psi))\ud{\psi},\qquad w(\psi)=4\pi \sin^2{\psi} \label{eq:energy}
\\
&&U(\psi,F(\psi),F'(\psi))
=\bigg\{
L\bigg[F'(\psi)^2+2\frac{\sin^2{F(\psi)}}{\sin^2{\psi}} \bigg]\nonumber\\
&&\qquad\qquad\qquad+\frac{2}{L}\bigg[F'(\psi)^2+\frac{1}{2}\frac{\sin^2{F(\psi)}}{\sin^2{\psi}}
\bigg] \frac{\sin^2{F(\psi)}}{\sin^2{\psi}}\bigg\}.\nonumber
\end{eqnarray}
As the unit of energy we chose $\e\f^{-1}/2$. Due to the spherical
symmetry imposed on solutions, the volume element on $\s{3}$ was
integrated over $\s{2}$ giving rise to the weight function
$w(\psi)=4\pi \sin^2{\psi}$. This function is important for
further considerations since it is natural to use it as the weight
function in the vector space of spherically symmetric functions
defined on $\s{3}$.

%% file: jmp_stg.tx
\section{Linear stability analysis. Some general introductory remarks}

Spherically symmetric, equivariant equilibria of the functional
(\ref{eq:lagdens}), that is (time independent) critical points of
energy functional (\ref{eq:energy}), may be stable or not in
dependence on whether they minimize the functional
$\mathcal{U}[F]$, which can be interpreted as potential energy. To
check stability of a map $F$ which is a solution of equation
$\delta\mathcal{U}[F]=0$  we compare its energy $\mathcal{U}[F]$
with energy of its deformation $\mathcal{U}[F+\varepsilon\xi]$,
where $\xi$ is a trial function satisfying some conditions (to be
stipulated later) and $\varepsilon$ is a small real number, which
is used as the parameter of the formal Taylor series expansion of
the functional $\mathcal{U}[F+\varepsilon\xi]$. In this paper we
are restricting ourselves to examining only the behaviour of the
term $\olarge{(\varepsilon^2)}$ of the expansion, by which it is
tantamount to the linear stability analysis of solutions $F$ (this
procedure is analogous to examining local extrema of ordinary
functions for which $f''\neq0$ at extremum). Our aim is to explain
bifurcations of solutions in our model and it suffices to examine
linear stability alone. This is also the reason we ignore in this
paper the cumbersome problem of marginal stability when a
perturbation does not change the term $\olarge{(\varepsilon^2)}$.
In what follows we will often skip the adjective 'linear' for
brevity. For an equilibrium solution to be stable we thus require
that in its vicinity the second variation of $\mathcal{U}[F]$ be
positive definite. The positive definiteness of
$\delta^2{\mathcal{U}[F]}$ is necessary and sufficient condition
that a critical solution was stable in the domain of spherically
symmetric perturbations. Nevertheless, it should be kept in mind
that a solution may turn out to be unstable if the constraining
conditions imposed on symmetries of perturbations are weakened.

Below a system will be said to be {\it energetically stable} if it
satisfies the above mentioned 'abstract' criterion of stability.
This is to distinguish it from the intuitively well understood
notion of {\it dynamical stability}. The last is being associated
with the evolution of a system in time. This evolution is
determined by a kinetic term of the total energy functional of the
system. Dynamical stability of a system in equilibrium is
understood as remaining arbitrary close to the equilibrium state
if only the perturbation is sufficiently small. The measure of the
perturbation may be determined by the amount of energy the
perturbation introduces into the system.

In appendix \ref{apdx:eigen} both these notions of stability are compared
one with another. This comparison was undertaken to see the
differences and whether they can affect the general statement that
a solution is linearly stable or not. To some extent these notions
turn out to be equivalent but some theorems on spectral analysis
are required when more than one mode of instability appears, since
the two approaches result in different Hilbert spaces of
perturbations (the difference comes from weight functions). This
will be elucidated in the following paragraphs.

\subsection{Formulation of the problem}

From now on one can forget about the genesis of energy functional
(\ref{eq:energy}) and consider it as defining an elliptic problem
for a spherically symmetric scalar field $F$ 'living' on the unit
three-sphere $\s{3}$. Then $L$ plays the role of a dimensionless
free parameter. Classical solutions of the model, that is the
critical points of functional (\ref{eq:energy}), satisfy equation
(\ref{eq:main}). This equation is necessary but not sufficient
condition for a solution $F$ to be a local minimum of the
functional. To see whether it is a minimum it suffices to check if
for a class of  perturbations $\varepsilon \xi$ the energy
$\mathcal{U}[F+\varepsilon \xi]$ is greater than $\mathcal{U}[F]$.
The perturbation $\xi$ can be normalized and $\varepsilon$ denotes
its (small in the sense of some scalar product) amplitude. If the
energy is increased for all perturbations within some given class
we say that solution $F$ is energetically stable with respect to
this class. For our purpose, among other things, we have to assume
this class to be composed of all perturbations which do not move
the considered fields $F$ away from the topological sector and do
not affect their spherical symmetry. The condition of symmetry for
perturbations allows us to reduce our problem and to treat it as a
one dimensional since one can consider the differential equation
(\ref{eq:main}) as such, regardless of its origin. Thus, one can
reformulate this problem as follows:
 in the domain $\psi\in[0,\pi]$ find
all regular solutions of (\ref{eq:main}) and check their
stability. To do this one may look for some variational principle
which reproduces the equation. But this may be done in many ways,
e.g. by inclusion of more spatial dimensions and time.

To give an example we can treat the integral (\ref{eq:energy}) as
a generalization of the energy functional of harmonic mappings
between three-spheres. Then to check energetical stability of
these generalized mappings we would follow \cite{biz} by choosing
simply $4\pi\sin^2{\psi}$ as the natural choice of the weight function
in space of spherically symmetric perturbations on $\s{3}$. On the
other hand, since originally we obtained (\ref{eq:main}) from the
Skyrme model on $\s{3}$, we would write down the wave equation for
spherically symmetric, time dependent and equivariant solutions
$F(\psi,t)$. Static solutions of this wave equation, the
equilibria, would also be solutions of (\ref{eq:main}). We would
then follow the stability analysis made for spherically symmetric
static solutions of the Einstein-Skyrme model (of self-gravitating
skyrmions) \cite{bizchmaj}. This would correspond to the stability
analysis in dynamical sense. Both the approaches  would lead to
the same Sturm-Liouville operator acting in the same space of
admissible functions of perturbations but endowed with different
scalar products -- thus we would obtain different Hilbert spaces.
The difference would come from the weight functions -- the first
method would give simply $w(\psi)$ while the second would give a
complicated solution-dependent weight function. To make things
worse, even in the case of one space dimension one can relate
solutions of (\ref{eq:main}) to equilibria of many $1+1$
dimensional field theories that differ one from another in kinetic
terms. In fact, to examine energetical stability of a field
$F(\psi)$ in one space dimension one has to introduce some weight
function for which, among other things, the most natural choice is
the standard volume element on the interval $[0,\pi]$. To this
weight function there corresponds some 1+1 field theory for which
$F(\psi)$ is an equilibrium solution, and whose kinetic term is
such, that the equation for eigenvibrations around equilibria is
the same as the Sturm-Liouville equation which originates from the
energetical stability analysis. In this case both the dynamical
and energetical stability analysis lead to the same results (to
see the correspondence between the weight function and the kinetic term
the interested reader is referred to appendix \ref{apdx:eigen}).

Consequently, it arises the problem of a definition of an
appropriate unitary space of admissible perturbations, that is the
class of functions which should be used to vary the functional
$\mathcal{U}$, and of the choice of a weight function used to
define scalar product in this space. From the above analysis it
follows the  arbitrariness in the choice of weight functions. The
consequence is the question how this arbitrariness affects the
spectrum and qualitative results of stability analysis. Can we
simplify the analysis of energetical stability of hedgehogs by
choosing simply $g=w$? (This choice of the weight function is due to
the 'simplicity and naturalness'). Put differently, we would
like to know whether the qualitative predictions of this analysis
are norm-invariant and, in particular, if the dynamical stability
analysis of time-dependent hedgehogs (which produces more
complicated Sturm-Liouville equation) would give the same
qualitative results?

As an aside, we remark also that from the point of view of the
functional (\ref{eq:lagdens}) the condition of spherical symmetry
for perturbations may turn out to be too stringent. In fact, a
solution which is stable under the stability criterion we assumed,
might turn out to be unstable in wider sense since, due to some
infinitesimally small nonspherical perturbation, it would evolve
to another configuration. With this reservation in mind we assume
the constraints of spherical symmetry on admissible perturbations
and, in addition to
 vanishing on the poles of the base $\s{3}$,
we assume them all to be pointwise continuously
differentiable. The last condition follows from the requirement
of continuity of the energy functional (\ref{eq:energy}).

\subsection{The Hessian and its weight function dependent spectrum }

The Hessian measures energies of excitations $\xi$ around a
classical solution $F$ and is quadratic with respect to
perturbations $\xi$. To avoid confusion between different
conventions, we define the Hessian as the coefficient
$\delta^2\mathcal{U}[F](\xi,\xi)$ in the series
$\mathcal{U}[F]+\delta\mathcal{U}[F](\xi)\varepsilon+
\delta^2\mathcal{U}[F](\xi,\xi)\varepsilon^2+\dots$ which is the
formal series expansion of the functional
$\mathcal{U}[F+\varepsilon\xi]$ with respect to the variable
$\varepsilon$, hence
\begin{equation}
\delta^2\mathcal{U}[F](\xi,\xi)=\int\limits_{0}^{\pi}w(\psi)\mathcal{F}
(\psi,\xi(\psi),\xi'(\psi))\ud{\psi}\label{eq:Hessian}
\end{equation}
where
\begin{eqnarray*} &&\mathcal{F}
(\psi,\xi,\xi')=
\left(L+\frac{2}{L}\frac{\sin^2{F}}{\sin^2{\psi}}\right)\xi'^2+
\frac{4}{L}\frac{\sin{2F}}{\sin^2{\psi}}F'\xi
\xi'\\&&\qquad\qquad\qquad+\left[\frac{2}{L}\left(1+2\cos{2F}\right)
\frac{\sin^2{F}}{\sin^4{\psi}}+
\frac{2\cos{2F}}{\sin^2{\psi}}\left(L+\frac{F'^2}{L}\right)\right]\xi^2.
\end{eqnarray*}
This is tantamount to defining an operator $\ho$ which acts in a
linear space of admissible functions $\mathcal{A}$ whose elements
are used as perturbations. If $\ket{\xi}\in \mathcal{A}$ then
$\ho$ is defined according to the formula
$\braket{\xi}{\ho}{\xi}:=\delta^2\mathcal{U}[F](\xi,\xi)$. For the
form $\delta^2\mathcal{U}[F](\xi,\xi)$ to be continuous we require
$\mathcal{A}$ to be composed of piecewise continuously
differentiable functions. Suppose that we have succeeded in
finding some countable and complete set of eigenvectors
$\ket{\xi_i}\in\ad$ to the corresponding eigenvalues $\lambda_i$
of the  operator $\ho$. To ensure this we require the form
$\braket{\eta}{\ho}{\xi}$ to be symmetric and the function space
$\mathcal{A}$ to be endowed with some scalar product
$\sprod{\cdot}{\cdot}{g}$ which, for any two vectors $\ket{\xi}$
and $\ket{\eta}$ from $\ad$, is defined by the integral
$$\sprod{\eta}{\xi}{g}:=\int\limits_{0}^{\pi}g\eta\xi\ud{\psi}.$$ The
weight function $g$ has to be positive for $\psi\in(0,\pi)$ and
fulfill some other conditions to ensure the existence of the above
integral for all elements of $\ad$. Now each perturbation
$\xi\in\ad$ can be uniquely decomposed in $\ad$ and thus written
equivalently in the form af an infinite  series
$\ket{\xi}=\sum_ic_i\ket{\xi_i}$. To excite the perturbation 
the amount of energy
$\braket{\xi}{\ho}{\xi}=\sum_i\lambda_ic_i^2\norm{\xi_i}{g}^2$ is
required. The representation of the function $\xi$ in the space
$\ad$ is unique and given by
$c_i=\sprod{\xi_i}{\xi}{g}\norm{\xi_i}{g}^{-2}$. Thus, to decide if
a solution $F$ is stable it suffices to check the positive
definiteness of the quadratic form $\braket{\xi}{\ho}{\xi}$, i.e.
to check if all eigenvalues are positive. From this it results the
following criterion of linear (energetical) stability:
\begin{dfn}
 For a solution of (\ref{eq:main}) to be linearly stable in
 the domain of spherically symmetric, pointwise continuously differentiable
  and
 vanishing on boundaries perturbations, it is necessary and
 sufficient that the resulting spectrum of
eigenvalues of Hessian (\ref{eq:Hessian}), that is of the second
variation of functional (\ref{eq:energy}), evaluated at this
solution, was positive. \label{dfn:stability}
\end{dfn}
Now the problem of a choice of the appropriate scalar product
arises. Anyway, before deciding this, we will find the operator
$\ho$
 and carry out its spectral decomposition assuming, temporarily,
  an arbitrary scalar product.
We assume that all  eigenvalues are enumerated in such a way they form a
nondecreasing sequence $\{\lambda_i\}$. Then it follows, analogously
as in the case of finite dimensional quadratic forms, that each
element $\lambda_i$ may be characterized by a process of
consecutive  minimizations \cite{courant}. The lowest eigenvalue
$\lambda_0$ is defined as the global minimum of the functional
$\hess\cdot\norm{\xi}{g}^{-2}$ in the domain $\ad$. By $\xi_0$ we
denote the minimizing $\xi$. The $n$'th eigenvalue is a minimum of
the functional $\hess\cdot\norm{\xi}{g}^{-2}$ under the assumption
that $\sprod{\xi}{\xi_k}{g}=0$, $k<n$ and the minimum is attained
by $\xi=\xi_n$. Applying this to the functional
$$\Lambda[\xi]:=\frac{\braket{\xi}{\ho}{\xi}}{\norm{\xi}{g}^2},
\quad \xi\in\ad$$ it is straightforward that, at least for the
global minimum $\xi=\xi_0$, the equation
\begin{equation}
0=\delta\Lambda[\xi]=2\frac{\braket{\delta\xi}{\ho}{\xi}-
\lambda\sprod{\delta\xi}{\xi}{g}}{\norm{\xi}{g}^2}, \quad
\mathrm{where}\quad \lambda:=\Lambda[\xi]
\label{eq:lambda}\end{equation} and consequently the equation
\begin{equation}
\frac{1}{2}\frac{\delta \left(\hess\right)}{\delta
\xi(\psi)}=\lambda g\xi\label{eq:eig}\end{equation}
or equivalently the equation
$$\frac{1}{2}\frac{\delta(w\mathcal{F})}{\delta
\xi(\psi)}=\lambda g(\psi)\xi(\psi),$$
 must
necessarily hold since $\norm{\xi}{g}>0$. The foregoing functional
derivative involves $F''(\psi)$ which, for solutions, can be
expressed by $F$ and $F'$ using (\ref{eq:main}), hence
\begin{eqnarray} &&-\frac{\ud}{\ud{\psi}}\left(4\pi
\left(L\sin^2{\psi}\label{eq:eigen}
 +\frac{2}{L}\sin^2{F(\psi)}\right)
\frac{\ud{\xi(\psi)}}{\ud{\psi}}
\right)\\ &&\qquad\qquad\qquad+\frac{8\pi
A(\psi)}{L\sin^2{\psi}+\frac{2}{L}\sin^2{F(\psi)}}\xi(\psi)=
\lambda g(\psi)\xi(\psi),\nonumber
\end{eqnarray}
where
\begin{eqnarray*}
 &&A(\psi)=\left(2\cos{2F(\psi)}-1\right)\sin^2{F(\psi)}
\\&&\qquad+L^2\left(1+
\frac{2}{L^4}\frac{\sin^4{F(\psi)}}{\sin^4{\psi}}\right)
\cos{2F(\psi)}\sin^2{\psi}\\ 
&&\qquad+F'(\psi)\sin{2\psi}\sin{2F(\psi)}+\frac{(F'(\psi))^2}{L^2}
\left(1-\cos{2F(\psi)}\left(1+L^2\sin^2{\psi}\right)\right).
\end{eqnarray*}
Equation (\ref{eq:eigen}) together with the class of admissible
functions $\ad$ satisfy the general requirements for the
Sturm-Liouville problem that the theorems proven in \cite{courant}
would be successfully utilized. It follows that the condition
$\sprod{\xi}{\xi_k}{g}=0$, $k<n$ in finding the other consecutive
minima $\lambda_n$ is sufficient that equation (\ref{eq:lambda})
would hold in general. It also follows from \cite{courant} that
the respective minimizing $\xi_n$ exist and are the same as
solutions of equation (\ref{eq:eigen}) which is the necessary
condition that the first variation of (\ref{eq:Hessian}) vanished.
 It also follows that the corresponding
eigenvalues $\lambda $ of (\ref{eq:eigen}) are the same as the
minima $\lambda_n$. Moreover
$\braket{\xi_i}{\ho}{\xi_j}=\lambda_i\norm{\xi_i}{g}\delta_{ij}$
and the denumerably infinite and, by construction, mutually
orthogonal set of eigenfunctions $\xi_i$ is complete in $\ad$.
Using the metric form $\sprod{\cdot}{\cdot}{g}$ we normalize these
functions to the unity. Thus, we have managed to construct the Hilbert
space $(\ad,\norm{\cdot}{g})$ in which $\ket{\xi_i}$ form an
orthonormal and complete base and in which the operator $\ho$ is consequently
represented by the formula
$$\sum\limits_{k=0}^{\infty}\nbra{g}{\xi_k}\lambda_k\nket{\xi_k}{g}.$$
The second order linear self-adjoint expression on the left  in
(\ref{eq:eigen}) is just the element $\braket{\psi}{\ho}{\xi}$.
  As an aside, we remark that in this way we were led
     straightforwardly to the differential operator $\ho$ which is
     naturally born as self-adjoint. Due to the boundary
     conditions imposed on elements of $\ad$ this operator is also hermitian.

Before we utilize the foregoing formalism to carry out the
stability analysis of solutions of equation (\ref{eq:main}), we
examine the problem of deciding which weight function should be
substituted in place of $g$. It is straightforward for spherically
symmetric functions defined on the unit three-sphere to normalize
them using the natural volume element $w(\psi)$, i.e. to
substitute $g=w$ into (\ref{eq:eigen}). In this paper we follow
this natural choice, especially, as then the comparison with the
results of harmonic maps between the three-spheres \cite{biz} may
be done. On the other hand, one would equally well argue for
another choice. In fact, the functional (\ref{eq:energy}) defines
a problem of finding extrema in the function space composed of
functions of one variable $\psi\in[0,\pi]$, therefore the natural
choice of the weight function would be simply $g(\psi)=1$. Another
possibility is given by the following physical argumentation. The weight
function should be chosen in such a way that the resulting
spectrum of the Hessian could be directly interpreted in terms of
frequencies of eigenvibrations of our system and, as such, should
be determined by the kinetic term alone. The appropriate $g$ was
constructed in appendix \ref{apdx:eigen}.

It seems there is no sufficiently strong criterion which would
determine the appropriate weight function. This signalizes that it
is rather quite arbitrary which weight function should be chosen
to normalize the function space of admissible perturbations, as
long as the axioms of  scalar product are satisfied. The situation
with the arbitrariness of the choice of 'g' resembles a
sort of gauge freedom. However a nontrivial change in the scalar product
inevitably affects the spectrum and the respective eigenfunctions.
Consequently, this changes the Hilbert space, so that
'observables' can not be  unitary transformed one to another.
Nonetheless, it is very plausible this is not a real obstacle and
may be successfully cured. Some arguments it is really the case
are given below.

From the minimizing properties of eigenvalues it trivially follows
the observation that the positive definiteness of the Hessian is
universal, i.e does not depend on the specific weight function
assumed. It is also  clear that the sign of the lowest eigenvalue
is universal as well. Thus, to answer the general question if a
solution is stable, it is quite arbitrary in which scalar product
the function space $\ad$ is endowed. This is in agreement with the
intuition that stability is not of dynamical origin. The problem
arises when one wants to know how many
 instabilities a particular solution possesses. The question is
whether the number of instabilities, i.e. if the number of
negative eigenvalues (which is always finite) is norm invariant?
Actually, a simple argument that the number of instability modes is
norm-invariant can be constructed.

It turns out that at a critical $L_c$, at which a new solution of
equation (\ref{eq:main}) appears or bifurcates from an already
existing one, the solution has an eigenvalue which vanishes.  Let
denote it by $\lambda_n$ where $n$ is the number of nodes of the
corresponding eigenfunction $\xi_n$. Since $\lambda_n=0$ it is
clear that this eigenfunction is universal at $L=L_c$, i.e. does
not depend on a specific weight function $g$ (to preserve the
asymptotics of solutions of (\ref{eq:eigen}) we may assume that
weight functions vanish on boundaries). Moreover, the number of
nodes of $\xi_n$, which is also the number of negative
eigenvalues, can not be affected by a continuous change of any
coefficient of the Sturm-Liouville equation (\ref{eq:eigen}). Due
to continuous dependence of eigenvalues on the coefficients a
negative eigenvalue must remain negative for all $L$ such that
$L_c<L<L_c+\varepsilon$ where $\varepsilon$ is positive and
sufficiently small. From the continuity and universality of
$\lambda_n(L_c)$ it also follows that the sign of $\lambda_n(L)$
must be universal for $0<L-L_c<\varepsilon$. Otherwise there would
exist such two weight functions $g(0)$ and $g(1)$ for which, respectively,
$\lambda_n(0,L)<0$ and $\lambda_n(1,L)>0$. Then for some
$\alpha\in(0,1)$ the equation (\ref{eq:eigen}) with $g=\alpha
g(1)+(1-\alpha)g(0)$ would have vanishing eigenvalue
$\lambda_n(\alpha,L)=0$ which, in turn, would not be universal, a
contradiction.

%% file: jmp_sts.tx
\section{Linear stability analysis of solutions $\sk{n}\ha{k}\sk{m}$.
Critical phenomena and spontaneously broken reflection symmetry}

To summarize, we have reduced the problem of examining the linear stability
of solutions of equation (\ref{eq:main}) to the Sturm-Liouville
problem for the differential operator (\ref{eq:eigen}) with the
condition that its eigenfunctions $\xi(\psi)$ must vanish  on
boundaries $\xi(0)=0=\xi(\pi)$. As thea weight function we chose
simply the volume element on $\s{3}$ -- $g(\psi)=4\pi\sin^2{\psi}$
-- as we gave some arguments that qualitative results of our
analysis can not be affected by this special choice. In this way
we also avoid the unnecessary complications introduced by the weight
function implied by the kinetic term of the Skyrme model on
$\s{3}$. Due to global regularity of solutions of equation
(\ref{eq:main}) the boundary points are regular singular points of
equation (\ref{eq:eigen}). Thus solutions of (\ref{eq:eigen}) in
the vicinity of boundaries can be written as generalized series.
From the secular equation of (\ref{eq:eigen}) it follows that the
first independent solution is a Taylor series for which $\xi$
vanishes at a boundary point and thus can be made vanishing on the second boundary point to satisfy the
requirements of admissibility. We also note the very useful for
numerical integrations fact, that (if not degenerate) the
eigenfunction
 $\xi_k$ corresponding to the 
eigenvalue $\lambda_k$ has exactly $k$ nodes within
the interval $(0,\pi)$ ($\xi_0$ has no nodes inside) and the
sequence  $\{\lambda_k\}$ is nondecreasing and divergent.
\subsection{$\ha{k}$ in the limit $L\to\infty$, harmonic maps}
Let $F$ be the solution $\ha{k}$ (then
 $F'$ is finite for all $\psi\in[0,\pi]$) and take the limit $L\rightarrow \infty$ in
 (\ref{eq:eigen}). In this way we reproduce the differential equation for linear
 perturbations of
harmonic maps $\has{k}$ of \bizon which was analyzed in \cite{biz}
(the author used the conformal variable $x=\ln{(\tan{(\psi/2)})}$)
$$-(\xi'\sin^2{\psi})' +2\cos{(2F)}\xi =\tilde{\lambda} \xi
\sin^2{\psi},\quad\tilde{\lambda}=\lim_{L\rightarrow \infty}
\frac{\lambda(L)}{L}.$$ The equation can be
 solved analytically for the identity solution, the harmonic map
 $\has{1}$ (the case of the vacuum $\ha{0}$ is trivial) then
$$\xi^1_l=\sin{\psi}C^2_{l}(\cos{\psi}), \quad \omega^1_l=l^2+4l-1,$$ where
$\xi^1_l$ (and respectively $\omega^1_l$) denote the $l$'th (where
$l=0,1,2,\dots$) eigenfunction (eigenvalue) around the regular
solution $\has{1}$ and $C^2_{l}$ are the Gegenbauer polynomials.
The map $\has{k}$ has exactly $k$ unstable ($SO(3)$-symmetric)
modes (for detailed information see \cite{biz}).

\subsection{Stability of the vacuum solution $\ha{0}$\label{par:h0}}
It is clear that the solution $\ha{0}$ ($\ha{0}=k\pi$, $k$ an
integer) whose energy is zero must be stable, since energy
integral (\ref{eq:energy}) is bounded from below by zero. Hence
any perturbation can only increase  energy, therefore the vacua
are stable. Nonetheless, we carry out the calculations since the
spectrum will be used to interpret spectra of perturbations around
$\sk{n}$ or $\sk{n}\sk{m}$. This is because skyrmionic solutions
tend pointwise (but not uniformly) to $\ha{0}$ in the limit
$L\to\infty$.

The substitutions $F(\psi)=k\pi$ and $\xi(\psi)\rightarrow
\sqrt{1-x^2}u(x)$, where $\psi= \arccos{(-x)}$, reduce equation
(\ref{eq:eigen}) to the eigenvalue problem
$$(1-x^2)u''(x)-5xu'(x)+\left(\frac{\lambda}{L}-3\right)u(x)=0,$$
which is the Gegenbauer equation for $u$. Due to the boundary
conditions imposed on $\xi$ the function $u$ must be bounded
everywhere in the closed interval [-1,1]. This is possible if
$\lambda= \lambda_l=L(l^2+4l+3)$, $l=0,1,\dots$.  Thus the
eigenvalue problem for $\ha{0}$  is solved and given by
$$\xi_l(\psi)=\frac{1}{\pi}\sqrt{\frac{2}{l^2+4l+3}}\sin{\psi}
C^2_l(\cos{\psi}), \quad \lambda_l=\lambda_l=L(l^2+4l+3), \quad
l=0,1,\dots,$$  where $C^{2}_l$ are the Gegenbauer polynomials.
Written explicitly, the few first normalized modes of the vacuum
reads respectively\begin{eqnarray*}
&&\frac{1}{\pi}\sqrt{\frac{2}{3}}\sin{\psi},\quad
-\frac{1}{\pi}\sin{2\psi}, \quad
\frac{2\sqrt{30}}{15\pi}\sin{\psi}\left(2+3\cos{2\psi}\right),\\
&&
\qquad\qquad\qquad\qquad\qquad\qquad\qquad\qquad\qquad
 -\frac{\sqrt{3}}{3\pi}\sin{2\psi}\left(1+4\cos{2\psi}\right)
.\end{eqnarray*}
Since all $\lambda_k>0$ the vacuum solution is stable as expected.
\subsection{On the instability of $\ha{1}$ which gives rise to
the appearance of $\sk{1}$ by a critical phenomenon at the
critical radius $L=\sqrt{2}$} \label{h1stab}
\label{par:appropriate} Substituting into (\ref{eq:eigen})
$F(\psi)=\psi$ we obtain the eigenvalue problem
\begin{equation}-(\sin^2{\psi}\xi'(\psi))'+
2\left(1-2\frac{L^2+1}{L^2+2}\sin^2{\psi}\right)\xi(\psi)=
\frac{L}{2+L^2}\sin^2{\psi}\lambda \xi(\psi)\label{eq:eigenh1}
\end{equation} which is solved analogously as before resulting with
the Gegenbauer equation
$$(1-x^2)u''(x)-5xu'(x)+\left(\frac{L^2-2}{L^2+2}+\frac{L}{L^2+2}\lambda
\right)u(x)=0$$ whose solutions are bounded everywhere in the closed
interval [-1,1] only if $\lambda=\lambda_l$ where
$$\lambda_l=\frac{2}{L}(l^2+4l+1)+L(l^2+4l-1).$$ The respective
eigenfunctions of (\ref{eq:eigenh1}) are the same as for $\ha{0}$.
(If we had used the weight function resulting from the kinetic
term of the Skyrme model we would have got
$\omega^2_k=L\lambda_l/(L^2+2)$ with $\lambda_l$ as above, see
appendix \ref{apdx:eigen}). In the exceptional cases of $\ha{0}$ and
$\ha{1}$ the respective eigenvectors do not change with $L$ for
the weight function assumed. It is no longer true in generic
case of  weight function. (The other solutions of (\ref{eq:main}),
known only numerically, change continuously as $L$ increases, thus
so do the eigenfunctions).

All $\lambda_l$ with $l\geq 1$ are positive for every $L$. The map
$\ha{1}$ is stable for $L<\sqrt{2}$ and consequently a minimum of
the functional (\ref{eq:energy}). For $L=1$ it also saturates the
Bogomolnyi bound $E=12\pi^2$. For $L>\sqrt{2}$ the map $\ha{1}$ is
no longer stable and by this the critical value $L=\sqrt{2}$ is
distinguished. When the threshold of stability is passed, the
new solution $\sk{1}$ appears. Due to the reflection symmetry of
equation (\ref{eq:main}): $F(\psi)\to\pi-F(\pi-\psi)$ it also
 appears the $1$-skyrmion $\askk{1}$ which is localized at 
the south pole for large $L$ (in what follows we will be using this notation 
for that $1$-skyrmion). Note that
$\sk{1}$ is not reflection symmetric unlike $\ha{1}$. Both
$\sk{1}$ and $\askk{1}$ bifurcate smoothly from $\ha{1}$ and exist
for $L>\sqrt{2}$ tending continuously to skyrmions localized at
the poles of the base three-sphere in the limit $L\to\infty$. In
figure \ref{gr:s1fmeig}
 the evolution of spectra of eigenvalues
 of (\ref{eq:eigen}) for $\sk{1}$ and $\ha{1}$ are shown.
\sglgr{h1fm_eig}{Evolution of eigenvalues for $\ha{1}$ (-------)
and $\sk{1}$ ($\cdots\cdots$). We use the definition
$\kappa^2:=2L^{-2}$ where $L$ is the radius of the base
three-sphere, and instead of eigenvalues $\lambda$ we show the
rescaled values $\lambda /L$ that are finite in the limit
$\kappa^2\to 0$. At the critical $\kappa^2=1$ ($L=\sqrt{2}$) the
lowest eigenvalue of $\ha{1}$ becomes positive, thus $\ha{1}$ is
stable for $\kappa^2>1$ and unstable for $\kappa^2<1$. Due to
marginal stability of $\ha{1}$ at the critical $\kappa^2=1$ the
skyrmionic solutions $\sk{1}$ and $\askk{1}$ appear bifurcating from
$\ha{1}$. The skyrmionic solutions are stable for $\kappa^2<1$ and
do not exist for $\kappa^2>1$.}{gr:s1fmeig}
Due to positive definiteness of the Hessian at $F=\sk{1}$ it
follows that the skyrmionic solution is stable (it possesses also
the lowest energy).

In what follows, we explain the numerical observations. In a
sufficiently small right neighbourhood of the critical $L$  it is
energetically preferable to excite  the mode $\xi_0$ on $\ha{1}$
since then the energy is diminished. It is also the only mode with
negative energy at our disposal. The exceptional role of the mode
$\xi_0$ is also reflected in the fact that it is the invariant
solution of the eigenvalue problem (\ref{eq:eigen}) at
$L=\sqrt{2}$ with respect to a change of the weight function. Thus
the solution has to play a nontrivial role in the vicinity of the
critical $L$.
\dblgr{h1fm_shp}{h1fm_en}{Shooting parameters -- $a$ -- (on the
left) and energies -- $E=\mathcal{U}[F]$ -- (on the right) for
$\ha{1}$ ($\cdots\cdots$) and $\sk{1}$ (-------) as functions of
$\kappa^2$ and $L$, respectively. We use the definition
$\kappa^2:=2L^{-2}$ where $L$ is the radius of the base
three-sphere. At $\kappa^2=1$ ($L=\sqrt{2}$) the skyrmionic
solution $\sk{1}$ bifurcates from $\ha{1}$ and exist for
$\kappa^2<1$. The curves of shooting parameters of $\sk{1}$ and of
$\ha{1}$ (for $\ha{1}$ $a(L)\equiv 1$) cross vertically each other
at the bifurcation point $\kappa^2=1$ while the curves of energies
(on the right) smoothly approach one to another. This
characteristic singular 'fork-like' structure in both diagrams
signalizes a critical phenomenon.}{gr:h1}
Close to the critical value $L=\sqrt{2}$ the plot of shooting
parameters of the map $\sk{1}$ (figure \ref{gr:h1}) is (at
confidence of 0.95) perfectly described by the fitting curve
\begin{eqnarray}&&L-\sqrt{2}=\delta+c(a-a_o)^b(1-c_1(a-a_o)+c_2(a-1)^2
+\dots),\label{eq:num}\\
&&\qquad a_o=0.9999997\pm4\times10^{-6},\quad b=2,0000007\pm0.0005,
\nonumber\\
&&\qquad c=0.2592695\pm5\times10^{-4},\quad \delta=-1.54\times10^{-9}\pm 10^{-9}\nonumber \\&&\qquad c_1=0.6001\pm0.02, \quad
c_2=0.50175\pm0.1.\nonumber
\end{eqnarray} This is the numerical confirmation of the
hypothesis that in the vicinity of the critical $L$ the shooting
parameter $a$ of the map $\sk{1}$ possesses the fractional
exponent behaviour \begin{equation}|a-1|\propto
(L-\sqrt{2})^{1/2}\label{eq:crit}\end{equation} characteristic for
critical phenomena. This behaviour is remarkable and it is
desirable that its analytical explanation was given. 
In what follows we will show that this result may be 
reproduced analytically together
with  the calculation of the proportionality factor in
(\ref{eq:crit}).

 In the vicinity of critical $L=\sqrt{2}$ we can
find three distinct formal Fourier series expansions of the function
$F(\psi)-\psi$ which depend on some small real parameter $\alpha$
$$F(\psi)-\psi\sim\sum\limits_{m=1}^{\infty}\alpha^ma_m(\alpha)\sin{(m\psi)},$$
where $a_m{\alpha}$ are functions to be determined and $\alpha\to0$ as
$L\to\sqrt{2}^+$.
Its shape guarantees that as $\alpha\to0$ then $F(\psi)\to\ha{1}$.
The other solutions $\sk{n}\sk{n}$ or $\sk{n}\ha{1}\sk{n}$, which
exist at $L<\sqrt{2}$, are unfortunately too remote from the
analytically known $\ha{1}$ to be examined in this way. The
coefficients $a_m(\alpha)$ are defined by the requirement that
$F(\psi)$ should satisfy equation (\ref{eq:main})  expanded in
$\alpha$. These coefficients may be found by a recurrence process
if we assume that they may be approximated by partial sums of the
form
$$a_m^{(N)}(\alpha)=\sum\limits_{k=0}^Na_{m,k}\alpha^{2k}, \qquad
\mathrm{with}\qquad \alpha^2:=L-\sqrt{2}>0.$$ Since the
eigenvalues of Hessian at $\ha{1}$ read $\xi_{2n}(\psi)
\propto\sum_{k=0}^{n} (2k+1)\sin{((2k+1)\psi)}$ and
$\xi_{2n+1}(\psi) \propto\sum_{k=1}^{n+1} 2k\sin{(2k\psi)}$ it is
clear that there exist
 formal (infinite) polynomials $b_m(\alpha)$ of the variable $\alpha^2$
 with $b_m(0)\neq0$
 such that the above Fourier series may be recast into the form
 $$F(\psi)\sim\psi+\sum\limits_{m=1}^{\infty}\alpha^{m}b_m(\alpha)\xi_{m-1}(\psi)$$
Before we proceed further, it should be stressed the fact that this procedure is complitely formal, since it presupposes a line of analytical properties that do not have to be valid. Therefore the procedure may lead to false conclusions, unless one proves its correctness.
The first of the three series is
simply zero and corresponds to the solution $\ha{1}$. To the order
of $\alpha^3$, to write only a few, the second is given by
\begin{eqnarray}&& F(\psi)\sim\psi+\left(\sqrt{\frac{30\sqrt{2}}{11}}-
\alpha^2\frac{29369}{638880}\sqrt{165\sqrt{2}}\right)\alpha\sin{\psi}
\nonumber \\&&\qquad\qquad +
\alpha^2\frac{9\sqrt{2}}{22}\sin{2\psi}+\alpha^3\frac{15\sqrt[4]{2}}{88}
\sin{3\psi}+o(\alpha^3)\label{eq:formal}\end{eqnarray} and the
third formal series may be derived from the last by the reflection
$F(\psi)\to \pi-F(\pi-\psi)$. The reader is referred to appendix
\ref{apdx:series} to see the expansion up to a higher order. For
$L<\sqrt{2}$ with $\alpha^2=\sqrt{2}-L$, a similar
procedure does not
give any real series apart from the one which is identically zero and
corresponds to $\ha{1}$. Expressed in the base of the eigenvectors
of the Hessian (\ref{eq:eigen}) evaluated at $\ha{1}$ the foregoing series reads
$$ F(\psi)\sim\psi+3\pi\sqrt{\frac{5\sqrt{2}}{11}}\cdot\alpha\xi_0(\psi)-
\frac{9\sqrt{2}}{22}\pi\alpha^2\xi_1(\psi)+o({\alpha^2}).$$ The
question arises whether the formal series (\ref{eq:formal}) has a
nonzero radius of convergence in the variable $\alpha$. It
requires a rigourous proof. Here we give only a naive argument
which makes it plausible that (\ref{eq:formal}) is really the
solution of (\ref{eq:main}) (the reader is also referred to
appendix \ref{apdx:series} to see how these coefficients behave). It was
proved in \cite{my} that finite energy solutions of equation
(\ref{eq:main}), in the vicinity of  the singular point $\psi=0$, are analytic functions of $\psi$. Thus, every such solution can be written as an infinite power series whose general form reads
$$F(\psi)=a\psi-\frac{a(a^2-1)(2L^2+a^2)}{15(L^2+2a^2)}\psi^3+
\olarge{(\psi^5)},\qquad a=F'(0).$$
The series contains only odd powers of $\psi$ (the series (\ref{eq:formal})
 as well) and the coefficients are rational (thus also analytic) functions of $a$ and $L$.
Substituting
 into the series  $L=\sqrt{2}+\alpha^2$ and  $a=F'(0)=1+c\alpha+\dots$ calculated
 from (\ref{eq:formal}) and next expanding in $\alpha$ we compare the result with
 (\ref{eq:formal}) expanded in $\psi$ (the comparison was carried out
 up to $\psi^{16}$ and
 $\alpha^{20}$ with positive result). Now we can compare the 'solution'
 (\ref{eq:formal}) with the numerical results for $\sk{1}$.

As one would expect this approach explains our numerical results.
The formal series predicts in the vicinity of $L>\sqrt{2}$ the
singular behaviour of shooting parameters of the map $\sk{1}$
$$|a-1|=\sqrt{\frac{30\sqrt{2}}{11}}\cdot(L-\sqrt{2})^{1/2}+
o((L-\sqrt{2})^{1/2}.$$ To compare with the numerically derived
curve (\ref{eq:num}) we rewrite the above formula into the form
$L-\sqrt{2}=0.2592725\dots\cdot(a-1)^2+o((a-1)^2))$. The comparison
is also a sort of accuracy test of the numerical
integrator used to solve equation (\ref{eq:main}). In spite of the fact
that  close
to the bifurcation points the numerics can not be very precise, this test
gives quite good results.

From the foregoing analysis it also follows that, to first order in
the critical parameter $\alpha$, the instability which causes
$\sk{1}$ (and $\askk{1}$) to appear by bifurcation from $\ha{1}$ at
$L=\sqrt{2}$, is due to the mode $\xi_0$ whose eigenvalue becomes
negative at $L=\sqrt{2}$. Thus the mode $\xi_0$ is in fact
distinguished and this is due to the nonanalyticity of the
function $a(L)$ at the critical $L=\sqrt{2}$. The function
$\sin{\psi}$ was used as a trial function to verify the
instability of the identity map in \cite{manton86} where the
Skyrme model on $\s{3}$ was proposed. Simultaneously, the authors
expressed the belief that with a better trial function it should
be possible to establish the instability of the identity map at
$L<\sqrt{2}$. It is clear that this must be performed by using trial
functions which are not spherically symmetric since, at is was shown before,
 $\ha{1}$ is a
local minimum of (\ref{eq:energy}) for $L<\sqrt{2}$ in the domain
of spherically symmetric functions.

 The formal solution $\sk{1}$ in the vicinity of critical $L$
(\ref{eq:formal})
 can be utilized to find a similar series expansion of energy  of $\sk{1}$.
To carry this out up to $\alpha^6$ we need to
 take into account contributions from the three first modes what
 gives (to see the expansion up to a higher order see 
appendix \ref{apdx:series})
$$\mathcal{U}[\sk{1}(\alpha)]=12\pi^2\left(\frac{3\sqrt{2}}{4}+
\frac{1}{4}\alpha^2-\frac{19\sqrt{2}}{88}\alpha^4+\frac{15627}{42592}
\alpha^6 +\olarge{(\alpha^8)}\right).$$ To compare with, we also
expand $\mathcal{U}[\ha{1}]$
$$\mathcal{U}[\ha{1}]=12\pi^2\left(\frac{3\sqrt{2}}{4}+
\frac{1}{4}\alpha^2+\frac{\sqrt{2}}{8}\alpha^4-
\frac{1}{8}\alpha^6+\olarge(\alpha^8)\right).$$ Subtracting we get
$(\mathcal{U}[\sk{1}(\alpha)]-\mathcal{U}[\ha{1}(\alpha)])/(12\pi^2)
\approx-0.482118\alpha^4+0.491900\alpha^6+\olarge(\alpha^8)$ which
is in agreement with the numerical result
$-0.482(1)\alpha^4+0.49(12)\alpha^6$. Thus, the energy of $\sk{1}$
behaves smoothly and its plot coalesce with this of $\ha{1}$ in a
characteristic cusp at the critical point $L=\sqrt{2}$ where both curves are
tangent one to another (figure \ref{gr:h1}).

The approach presented here has the disadvantage that one can not
be sure that the series (\ref{eq:formal}) is really a solution of
(\ref{eq:main}) unless the appropriate proof is known
(nevertheless the numerics was quantitatively and qualitatively
explained). Therefore we give qualitative and mathematically correct
explanation (nonetheless quantitatively inaccurate). The evolution
of the profile $\sk{1}$ as $L$ grows resembles very much the
conformal deformation of the map $\ha{1}$ and it is well seen if
instead of $\psi$ one uses the conformal variable
$x=\ln{\tan{(\psi/2)}}$ \cite{my}. It should be noted that the
first time the conformal deformation of the identity solution was
examined  was by Manton in \cite{manton87} and it was in
connection with the stability analysis of the identity solution.
(our approach and motivation here is quite different). 
Here we compare the instability result
with the conformal deformation of the identity to construct a
function which would have a similar singular behaviour at the
critical $L$. The conformally deformed $\ha{1}$ (denoted by
$\ha{1}^{\beta}$) can also be decomposed in the base of
eigenfunctions $\xi_k$ of the Hessian at $F=\ha{1}$
\begin{eqnarray*} 
2\arctan{\left(e^{\beta} \tan{\frac{\psi}{2}}\right)}=
    \psi+\frac{\pi\sqrt{6}}{2}\left(1-
    \frac{1}{9}\beta^2+o(\beta^2)\right)\beta\xi_0\\
-\frac{\pi}{4}\left
    (1-\frac{11}{48}\beta^2+o(\beta^2)\right)\beta^2\xi_1+\dots
\end{eqnarray*}
The energy of $\ha{1}^{\beta}$ can be found
exactly and is given by
$$\mathcal{U}[\ha{1}^{\beta}]=6\pi^2\left(\frac{\cosh{\beta}}{L}+
\frac{L}{\cosh^2{(\beta/2)}}\right).$$ To chose the conformal parameter as
good as possible we can define it by the requirement that this energy, i.e.
 the function $f(\beta)=\mathcal{U}[\ha{1}^{\beta}]$, was at minimum.
For $L<\sqrt{2}$
 $f'(\beta)>0$ thus $f(\beta)$ attains its minimal value at $\beta=0$
 which is the energy of the map $\ha{1}$. For $L>\sqrt{2}$ the equation
 $f'(\beta)=0$ is solved also for some $\beta(L)\neq0$ which corresponds to 
the energy
 $$\mathcal{U}[\ha{1}^{\beta(L)}]=12\pi^2\left(\sqrt{2}-\frac{1}{2L}
 \right)<\mathcal{U}[\ha{1}], \qquad L>\sqrt{2}.$$ The function $\beta(L)$
 has the analogous interpretation as $a(L)-1$ for $\sk{1}$ and also
 possesses
 the characteristic critical behaviour $\beta(L)\approx\sqrt(L-\sqrt{2})$
 in the vicinity of critical $L=\sqrt{2}$. Thus again we have 
constructed a function
 which is singular at $L=\sqrt{2}$ and the corresponding profile $F(\psi)$
 (which is not solution of (\ref{eq:main})) has
 perfectly smooth energy. In the limit $L\to\infty$ this energy is finite
 and close to the energy of $\sk{1}$ for which, in this limit, we have $12\pi^2
 \cdot 1.23145$. Moreover. in this limit $F'(0)$ of $\ha{1}^{\beta(L)}$
 has the analogous behaviour as the shooting parameter $a_0$ of $\sk{1}$, that
 is $F'(0)\sim L$. The shooting parameter $a_{\pi}$ of $\sk{1}$ at the south pole   behaves
 like $L^{-2}$ for large $L$ while for $\ha{1}^{\beta(L)}$ it behaves like $L^{-1}$.
 Anyway, in this limit, the solution $\sk{1}$ is localized at the pole
 $\psi=0$ thus $\ha{1}^{\beta(L)}$, in a sense, well shows qualitative
 properties of $\sk{1}$ and possesses the analogous behaviour at the critical point. This
 enables us to qualitatively understand the critical phenomenon.

As an aside, we remark the fact, that the critical behaviour is
observed neither in the Skyrme model on flat space nor in the
model of harmonic maps between three-spheres when free parameters
of these models -- the characteristic size of a soliton in the
first case, or the radius of the base three-sphere in the second
case -- are changed. This shows that by coupling together two field
theories we may produce new phenomena that are absent in the
decoupled case. Dimensional parameters introduced by different
models determine then some numbers whose value is usually crucial
to the existence of different types of solutions. This fact is
known when nonlinear fields are coupled to gravity
\cite{bizchmaj}.

In what follows, we shall say that $F(\psi)$ has parity $+1$ or 
$-1$ if, respectively, 
$$f(\phi)=f(-\phi)\quad \mathrm{or}\quad
f(\phi)=-f(-\phi),$$
where
$$f(\phi):=F(\pi/2+\phi)-\frac{F(0)+F(\pi)}{2}, \quad \phi\in[-\pi/2,+\pi/2].$$
The solution
$\sk{1}$ has indefinite parity, unlike $\ha{1}$, from which it
bifurcates. This phenomenon can be easily explained by analyzing, 
in the vicinity of $L=\sqrt{2}$, 
the decomposition of $\sk{1}$ in the base of eigenfunctions of the Hessian
evaluated at $\ha{1}$. This is another remarkable feature of the critical
phenomenon in the Skyrme model on $\s{3}$. We postpone this
explanation until the last paragraph where quite analogous behaviour is
observed when other solutions appear by bifurcations.
\section{Linear stability analysis of solutions found numerically}

 To classify solutions of equation (\ref{eq:main}) the notation
 $\sk{n}\ha{p}\sk{m}$ was introduced \cite{my} to stress the fact
 that in the limit $L\to\infty$ a given solution  tends to a limiting
 configuration composed of a harmonic map $\has{p}$ of \bizon \cite{biz}
 localized in between the poles of the base three-sphere, to which
  the flat space skyrmions are attached at the poles, respectively,
 the $n$-skyrmion is localized
 at the north pole $(\psi=0)$ and the $m$-skyrmion is localized at the south
 pole. This configuration is characterized by
 a total topological charge $\q$ which is equal to $n+m$ or to $n+m\pm
 1$,
 respectively, if the solution contains an even or an odd harmonic map. The symbol $\sk{n}\ha{p}\sk{m}$ is used also to denote a
 class of solutions that can be transformed one to another by reflections
 which are the symmetries of equation (\ref{eq:main}). We assumed also the
 conventions $\sk{m}\equiv\as{-m}$, $\q(\ha{2k-1})=-\q(\ah{2k-1})$,
 $n\geq0$, where $m$ is such, that $0\leq\q\leq2n$ ($0\leq\q\leq2n+1$) for
 $p$ even (odd). To solutions that between the poles tend
 in the limit $L\to \infty$ to the vacua we will refer as pure
 skyrmionic solutions. To number eigenvalues of equation (\ref{eq:eigen})
 we use the convention that they are labelled by the
number of nodes of the respective eigenfunctions to which 
they corresponds, in particular the
lowest eigenvalue is denoted by $\lambda_0$.
\subsection{Nonsymmetric solutions possess constant number of modes
of instability. Pure skyrmionic solutions are stable} Regardless
of whether a solution of (\ref{eq:main}) appears together with its
companion (as in the special case where $\sk{3}$ appears together
with $\sk{2}\ha{1}$, figure \ref{gr:s1h3}) or bifurcates from an
already existing one (e.g $\sk{1}$ bifurcates from $\ha{1}$), this is
the rule, that this occurs always in conjunction with vanishing of
some eigenvalue of the Hessian. This paragraph summarizes the
observations made for
 nonsymmetric solutions, i.e. for solutions which have indefinite parity.
\dblgr{s3fm_eig}{s1h3fm_eig}{Evolution of eigenvalues (instead of
eigenvalues $\lambda$ here are shown the values $\lambda/L$ as
functions of $\kappa^2:=2/L^2$) for nonsymmetric solutions
$\sk{3}$ (-------) and $\sk{2}\ha{1}$ ($\cdots\cdots$), on the
left and for $\sk{1}\ha{3}$ (-------), $\sk{2}\ah{2}$
($\cdots\cdots$) on the right. }{gr:s1h3}
There exist two cases.
%
\begin{enumerate} \item  \label{pt2} For a
fixed triad of integers $(n,k,m)$ a solution
$\sk{n}\ah{2k-2}\as{n-m}$ appears together with
$\sk{n-1}\ha{2k-1}\as{n-m}$. The indices are such that $k\geq1$,
$n\geq1$ and $m$ is arbitrary as far as the inequality
$1\leq\q\leq2n-2$ is satisfied. Then the eigenvalue
$\lambda_{2k-2}^{nkm}$ vanishes at a characteristic for each triad
$(n,k,m)$ critical value $L=L^*_{nkm}$ (both the solutions have
the same spectrum only at $L^*_{nkm}$). The $L^*_{nkm}$ is the
same value of $L$ at which this pair appears, and respectively,
for $\sk{n}\ah{2k-2}\as{n-m}$ or $\sk{n-1}\ha{2k-1}\as{n-m}$ the
eigenvalue $\lambda_{2k-2}^{nkm}$ is positive or negative as
$L>L^*_{nkm}$. For $L>L^*_{nkm}$ the map $\sk{n}\ah{2k-2}\as{n-m}$
has $2k-2$ and $\sk{n-1}\ha{2k-1}\as{n-m}$ has $2k-1$ modes of
instability. The number of instabilities is the same as the number
of instabilities of the respective harmonic maps $\has{2k-2}$ or
$\has{2k-1}$ to which these solutions tend pointwise in between
the poles as $L\to\infty$. In figure \ref{gr:s1h3} are shown
spectra of the pair $\sk{2}\ah{2}$ and $\sk{1}\ha{3}$
corresponding to the triad $(2,2,2)$, and of the pair $\sk{3}$ and
$\sk{2}\ha{1}$ corresponding to $(3,1,3)$.
\item  For a fixed triad of integers $(n,k,m)$ the solution
$\sk{n}\ah{2k-1}\as{n-m}$ appears together with
$\sk{n-1}\ha{2k}\as{n-m}$. The indices are such that $k\geq1$,
$n\geq1$ and $m$ is such that the condition $1\leq\q\leq2n-2$
holds. Then $\lambda_{2k-1}=0$ at a characteristic for each pair
its own critical value of $L$ at which this pair appears.
Similarly, as before, these solutions possess constant number of
instabilities and inherit them from the respective harmonic maps.
\end{enumerate}
Summing up, each nonsymmetric solution possesses  a  constant
number of instabilities. This number is equal to the index of a
harmonic map to which the solution tend pointwise in the limit
$L\to\infty$. Note also that within a given pair its members
differ by one in the number of instabilities. There also exist a
subclass of (\ref{pt2}) which is composed of pairs containing 
pure skyrmionic
solutions. The last are characterized by the property that they
are stable, i.e. pure skyrmionic solutions are always stable.

Numerics also show (and this is true for all solutions, and can be
easily proved by examining equation (\ref{eq:eigen})) that the
spectra of eigenvalues of equation (\ref{eq:eigen}) with the
weight function $g(\psi)=4\pi\sin^2{\psi}$, in the limit
$L\to\infty$, tend (if divided by $L$) to the spectra of
eigenvalues of the harmonic maps between three-spheres \cite{biz}.
 This correspondence was the main motivation
for the choice of this special $g$. Of course, there exist a more
general class of weight functions for which such limiting
behaviour would be observed (it should be clear that, due to the
properties of our solutions,  it would happen e.g. for the $g$
which originates from the kinetic term (\ref{eq:skyrmkinetic}) and
from definition (\ref{eq:kineticnorm})). The same rescaling of
energies (\ref{eq:energy}) reproduces in this limit energies of
harmonic maps \cite{my}.

\subsection{Symmetric solutions $\sk{n}\sk{n}$, $n\geq1$
(of parity $(-1)$)} This subclass of pure skyrmionic solutions is
distinguished by the fact, that all the solutions exist for all
values of $L$. These solutions have parity $(-1)$ and positive
definite spectra (figure \ref{gr:s1s1}). Due to the positive
definiteness we conclude that the solutions  $\sk{n}\sk{n}$ are
stable for all $L$.
\dblgr{s1s1_shp}{s1s1_eig}{Evolution of shooting parameters (on
the left) and of eigenvalues (on the right) for the stable
solution $\sk{1}\sk{1}$. Instead of eigenvalues here are given the
rescaled values $\lambda/L$, also here is introduced the variable
$\kappa^2=2/L^2$. Solutions $\sk{n}\sk{n}$ with $n\geq2$ poses
qualitatively similar and positive spectrum of eigenvalues, thus
are also stable for all $L$}{gr:s1s1}
Apart from the poles of the base three-sphere these solutions tend
pointwise to the vacua. Also for all $n$ the spectra of
eigenvalues tend
 to the universal spectrum of $\ha{0}$ which was calculated in \S\ref{par:h0}.

\subsection{Symmetric solutions $\sk{n}\ha{1}\sk{n}$, $n\geq1$
(of parity $(-1)$)} These solutions exist for all $L$ and are
stable for sufficiently small $L<L^*_n$ where $\{L^*_n\}$ is
a monotonically increasing sequence of critical values of $L$.
\dblgr{s1h1s1fm_shp}{s1h1s1fm_eig}{Evolution of shooting
parameters (on the left) and of eigenvalues (on the right) for
solutions $\sk{1}\ha{1}\sk{1}$ (-------) and $\sk{2}\sk{1}$
$(\cdots\cdots)$. Instead of eigenvalues here are given the values
$\lambda /L$. Here is also introduced the variable
$\kappa^2:=2/L^2$. Solutions $\sk{n}\ha{1}\sk{n}$ with $n\geq2$
possess qualitatively similar in behaviour spectrum, that is for
some critical $L$ the lowest eigenvalue of $\sk{n}\ha{1}\sk{n}$
becomes negative and this is associated with the bifurcation of
the new stable solution $\sk{n+1}\sk{n}$.}{gr:s1h1s1}
For a fixed $n\geq1$ (this is also true for $n=0$, i.e. $\ha{1}$
\& $\sk{1}$ -- the case which was already analysed) the transition
between the stable and the unstable stage of the solution
$\sk{n}\ha{1}\sk{n}$ is accompanied by the appearance of the pure
skyrmionic solution $\sk{n+1}\sk{n}$, together with
$\sk{n}\sk{n+1}$, which are not reflection symmetric solutions and
have positive definite Hessian -- thus are stable. As an example
in figure \ref{gr:s1h1s1} are  shown characteristics of the
solution $\sk{1}\ha{1}\sk{1}$. For $L>L^*_n$ the solution
$\sk{n}\ha{1}\sk{n}$ inherits its single instability from the
harmonic map $\has{1}$ to which, between the poles, it pointwise
tends as $L\to\infty$. For all $n$ the limiting spectrum is
reproduced by eigenvalues of $\ha{1}$.

The critical behaviour of solutions $\sk{n}\ha{1}\sk{n}$ is quite
analogous to the behaviour of the identity $\ha{1}$. In
particular, due to the transition in instability at the critical
$L^*_n$ new solutions $\sk{n+1}\sk{n}$ and its reflection
$\sk{n}\sk{n+1}$, both with spontaneously broken reflection
symmetry, appear by bifurcation from $\sk{n}\ha{1}\sk{n}$. In more detail it is described in
the next paragraph.
\subsection[Symmetric solutions $\sk{n}\ha{2k+1}\sk{n}$ of parity
$(-1)$, $\sk{n}\ha{2k}\as{n}$ of parity $(+1)$, and accompanying
solutions, $n\geq0$, $k\geq1$]{Symmetric solutions
$\sk{n}\ha{2k+1}\sk{n}$ of parity $(-1)$, $\sk{n}\ha{2k}\as{n}$ of
parity $(+1)$, and accompanying solutions, $n\geq0$, $k\geq1$,
(thus also stable pure skyrmionic solutions $\sk{n}\as{n}$ of
parity $(+1)$ and unstable pure harmonic solutions of parity
$(\pm1)$)}  The analogous critical phenomenon, by which the
solution $\sk{1}$ was born by instability of $\ha{1}$, we come
across by observing the evolution of the reflection symmetric
solutions $\sk{n}\ha{2k-1}\sk{n}$ and $\sk{n}\ha{2k}\as{n}$. This
instability is responsible for the appearance of new solutions by
bifurcations. Unfortunately, unlike in the case of $\sk{1}$, 
these instabilities can not be analyzed here
analytically since all solutions but $\ha{1}(=\sk{0}\ha{1}\sk{0})$
are known only numerically.

A solution $\sk{n}\ha{2k}\as{n}$, $n\geq0$, $k\geq1$ (with
eigenvalues $\lambda_l=:\alpha_{n,l}$) appears together with
$\sk{n+1}\ah{2k-2}\as{n+1}$ ($\lambda_l=:\beta_{n,l}$) at some
characteristic $L:=L_{n,2k}$ when
$\alpha_{n,2k-2}=0=\beta_{2k-2}$. Both solutions are of  parity
$+1$. (If $k=1$ the last solution is simply $\sk{n+1}\as{n+1}$
thus is pure skyrmionic solution). For $L>L_{n,2k}$ the solution
$\sk{n+1}\ah{2k-2}\as{n+1}$ has $2k-2$ modes of instability, which
is the same as the number of instabilities of the harmonic map
$\has{2k-2}$. Moreover, $\alpha_{n,l}<\beta_{n,l}$ for
$L>L_{n,2k}$. For $L_{n,2k}<L<L^*_{n,2k}$ the eigenvalue
$\alpha_{n,2k-1}$ is positive, tends to $0$ as $L\to L^*_{n,2k}$
and is negative for $L>L^*_{n,2k}$. Thus,  for $L>L^*_{n,2k}$
$\sk{n}\ha{2k}\as{n}$ has $2k$ modes of instability -- the same as
the number of modes of the harmonic map $\has{2k}$. At the
critical $L^*_{n,2k}$ the solution $\sk{n+1}\ah{2k-1}\as{n}$ and,
due to reflection symmetry, the solution $\sk{n}\ha{2k-1}\as{n+1}$
appear by bifurcation from the solution $\sk{n}\ha{2k}\as{n}$. As
an example in figure \ref{gr:h2} are presented parameters of the
solutions $\ha{2}$, $\sk{1}\sk{1}$ and $\sk{1}\ah{1}$.
\dblgr{h2fm_shp}{h2fm_eig}{The evolution of shooting parameters
and eigenvalues for solutions $\ha{2}$ ({\bf -------}),
$\sk{1}\sk{1}$ ($\cdots\cdots$), and $\sk{1}\ah{1}$ (with its
partner $\ha{1}\as{1}$ ) (-------). This is a particular case of
more general situation when $\sk{n}\ha{2k}\as{n}$ ($n\geq0$,
$k\geq1$), together with $\sk{n+1}\ah{2k-2}\as{n+1}$ appear and,
due to the transition in the number of instabilities of
$\sk{n}\ha{2k}\as{n}$, the solution $\sk{n+1}\ah{2k-1}\as{n}$
appears by the process of bifurcation. All the solutions have
vanishing topological charge.}{gr:h2}
The solutions $\sk{n}\ha{2k}\as{n}$, $n\geq0$, $k\geq1$ and
$\sk{n+1}\ah{2k-2}\as{n+1}$ differ by $2$ in the number of
negative eigenvalues if $L$ is sufficiently large. This is the
first solution which possesses $2$ modes of instabilities more than
the second solution, and which
'bears' the new solution.
 The new solution has $2k-1$ modes of instability
(the same as the map $\has{2k-1}$) and is nonsymmetric, unlike the
one from which it bifurcates.

The solution $\sk{n}\ha{2k+1}\sk{n}$, with eigenvalues $\lambda_l=
:\alpha'_{n,l}$, appears together with
$\sk{n+1}\ah{2k-1}\sk{n+1}$, whose eigenvalues are
$\lambda_l:=\beta'_{n,l}$, at some characteristic $L=L_{n,2k+1}$
at which $\alpha'_{n,2k-1}=0=\beta'_{n,2k-1}$ (figure
\ref{gr:h3}).
\dblgr{h3fm_eig}{s1h3s1fm_eig}{Evolution of eigenvalues for
solutions $\ha{3}$ ({\bf -------}), $\sk{1}\ah{1}\sk{1}$
($\cdots\cdots$), and $\sk{1}\ah{2}$ ($\ha{2}\as{1}$) (-------) on
the left, and to compare with, the same characteristics for
$\sk{1}\ha{3}\sk{1}$ ({\bf
-------}), $\sk{2}\ah{1}\sk{2}$ ($\cdots\cdots$), and
$\sk{2}\ah{2}\sk{1}$. Instead of eigenvalues $\lambda$, are shown
$\lambda/L$, here is used also the variable $\kappa^2=2/L^2$. Both
figures are misleadingly similar, note the difference between the
critical values of $\kappa^2$ and tiny differences between the
critical values of eigenvalues in both figures. Because of this
one would have the impression that the presence of skyrmions has
not very crucial influence on the stability, but it suffices to
compare this with the spectrum of $\sk{1}\ha{3}$ (figure
\ref{gr:s1h3}) -- which is not a symmetric solution -- to see this
is not the case. These solutions are the particular case of the
more general situation when $\sk{n}\ha{2k+1}\sk{n}$ ($n\geq0$,
$k\geq1$) together with $\sk{n+1}\ah{2k-1}\sk{n+1}$ appear and,
due to the transition in the number of instabilities of
$\sk{n}\ha{2k+1}\sk{n}$ the solutions $\sk{n+1}\ah{2k}\sk{n}$ (and
$\sk{n}\ha{2k}\sk{n+1}$) appear by the process of bifurcation.
These solutions are topologically nontrivial $\q=2n+1$. }{gr:h3}
These solutions have also definite parity, i.e.
$(-1)$. For $L>L_{n,2k+1}$ the solution
$\sk{n+1}\ah{2k-1}\sk{n+1}$ has $2k-1$ modes of instability, which
is the same as the number of instabilities of the harmonic map
$\has{2k-1}$. Moreover $\alpha'_{n,l}<\beta'_{n,l}$ for
$L>L_{n,2k+1}$. For $L_{n,2k+1}<L<L^*_{n,2k+1}$ the eigenvalue
$\alpha'_{n,2k}$ is positive, tends to $0$ as $L\to L^*_{n,2k+1}$
and is negative for $L>L^*_{n,2k+1}$. Thus,  for $L>L^*_{n,2k+1}$
the solution $\sk{n}\ha{2k+1}\sk{n}$ has $2k+1$ modes of
instability -- the same as the number of instabilities of the
harmonic map $\has{2k+1}$. This is the critical value
$L^*_{n,2k+1}$ at which the solution $\sk{n+1}\ah{2k}\sk{n}$ (and
 $\sk{n}\ha{2k}\sk{n+1}$)
appear by bifurcation from $\sk{n}\ha{2k+1}\sk{n}$, i.e. from this
one of the original pair, which has $2$ more, than the second,
modes of instability. The new solutions have constant number of
$2k$ modes of instability (the same as $\has{2k}$) and have
spontaneously broken parity.
\section{The mechanism of the spontaneous breaking of the reflection symmetry}
The instabilities described above are closely related to the
breaking of reflection symmetry. As the rule we state that
solutions which bifurcate from symmetric solutions are not
symmetric and that only the solutions which are symmetric turn out
'to be able to give birth' to  new solutions. The special class of hedgehog
solutions is described by the reflection symmetric equation
(\ref{eq:main}), but some solutions possess no definite parity.
This phenomenon  is known as
 the spontaneous breaking of reflection symmetry or of parity.
 In the case of the Skyrme model on $\s{3}$ the
 following reflection symmetries of equation
(\ref{eq:main}) are being spontaneously broken: $F(x)\to F(-x)$
and $F(x)\to\q\pi-F(-x)$ (we assume $\lim_{x\to-\infty}F(x)=0$).
 We used here the variable
$x=\ln{\tan{(\psi/2)}}$. The second symmetry is a composition of
several primitive symmetries. Because of this we ascribe a
definite parity $+1$ or $-1$ to a solution which is invariant with
respect to the first or with respect to the second transformation.
By examining the process of the appearance of $\sk{1}$ by
bifurcation from $\ha{1}$, which was performed analytically, the
phenomenon of spontaneous breaking of the reflection symmetry can
be understood. By comparison with numerical data of other
solutions we can generalize our observations. As it will be shown,
the explanation of the symmetry breaking can be successfully
carried out with the use of the modes whose eigenvalues vanish at
some critical radii. This construction follows straightforwardly
from the analytical approach of  \S\ref{h1stab}.  Here it should
be noted that the first time the wording 'spontaneous breaking of
full rotational symmetry' in the case of Skyrme model on $\s{3}$
was used, was in the paper \cite{manton86}.

By a critical mode we will call the mode to which it correspond
the eigenvalue which vanishes at some critical $L=L^*$. We
observed that for $L<L^*$ this eigenvalue is positive and negative
for $L>L^*$ in all cases. It is worth to note the fact that a
critical mode has always definite parity at least for $L=L^*$.
This is because the solutions that 'give birth' to other solutions by
bifurcations have definite parity, and we observed that only such
solutions possess critical modes. The symmetry of the
Sturm-Liouville equation (\ref{eq:eigen}), which determines these
modes for such solutions, can be broken only by the weight
function which can be arbitrary. In fact, the vanishing of an
eigenvalue means that the corresponding eigenfunction is not
affected by the weight function, and by this is intimately
connected with the Hessian alone and consequently with the
equation (\ref{eq:main}). Thus, by analogy with \S\ref{h1stab} and
by the observation of similar behaviour of shooting parameters in
the vicinity of critical points $L^*_{n,p}$, at which new
solutions appear by bifurcations, we can write down the formulas
that describe, in the first approximation, the shooting parameters of
solutions $\sk{n+1}\ah{2k-1}\as{n}$ and $\sk{n}\ha{2k-1}\as{n+1}$
(which bifurcate from $\sk{n}\ha{2k}\as{n}$) or
$\sk{n+1}\ah{2k}\sk{n}$ and $\sk{n}\ha{2k}\sk{n+1}$ (which
bifurcate from $\sk{n}\ha{2k+1}\sk{n}$). For sufficiently small
$L>L^*_{n,p}$ these formulas read respectively
$\sk{n}\ha{2k}\as{n}\pm\gamma_{n,2k}\xi_{n,2k-1}$ or
$\sk{n}\ha{2k+1}\sk{n}\pm\gamma_{n,2k+1}\xi_{n,2k}$ where
$\xi_{n,p}$ is the critical mode and the coefficient
$\gamma_{n,p}$ shows the characteristic for critical phenomena
rational power dependence on $(L-L^*_{n,p})$ (the exponent was
calculated in \S(\ref{h1stab}) to be exactly $1/2$ and with very
good accuracy this value was also derived by fitting curves to
numerical data for several solutions). Note also, that the mode
$\xi_{n,2k-1}$ has an odd number of nodal points (hence the parity
$-1$) while the solution $\sk{n}\ha{2k}\as{n}$ has parity $+1$.
Analogously $\xi_{n,2k}$ has parity $+1$ which is opposite to the
parity of $\sk{n}\ha{2k+1}\sk{n}$. Thus, the solutions which
appear by bifurcation process have indefinite parity, i.e. have
spontaneously broken reflection symmetry. In the first
approximation, it is due to the excitation of the critical mode
which has the opposite parity with respect to the parity of the
excited reflection symmetric solution. This occurs while the
qualitative transition takes place, when the reflection symmetric
solution undergoes increase by $1$ in its number of instabilities.
In other words, one can say that at the critical point, in a
sense, it costs no energy of the excitation of the critical mode.
This is always inevitably connected with the process of
spontaneous breaking of reflection symmetry.

%% file: jmp_con.tx
In this paper the linear stability analysis of the whole spectrum
of the Skyrme's hedgehogs on the three-sphere was carried out.We considered only spherically symmetric perturbations. The
most remarkable it proved the influence of instability modes on the
existence of a class of solutions
 with spontaneously broken parity and which appeared by a critical phenomenon.
Also some general remarks on the construction of the Hilbert space
of admissible perturbations were presented together with the
resulting problem of a choice of a norm and its influence on the
resulting spectrum of perturbations.

The main lesson of this paper is that for $L$ sufficiently large ($L$ is
interpreted as the radius of the base three-sphere)
 the number of instabilities
of solutions $\sk{n}\ha{k}\as{m}$ is the same as the index of the
harmonic map $\has{k}$ to which (apart from the poles) the
solutions tend pointwise as the radius grows. The appearance of
new solutions is always accompanied by the vanishing of some
eigenvalue of the spectrum of the Hessian. In particular,
pure skyrmionic solutions $\sk{n}\sk{m}$ localized at the
poles are always linearly stable. The number of instabilities of
nonsymmetric solutions is independent on $L$. There
also exist a class of reflection symmetric solutions whose number
of negative modes increase by one as $L$ grows. Due to the
qualitative transition new solutions with broken reflection
symmetry appear by bifurcations and this process possesses many
characteristics of a critical phenomenon. In particular, the
identity solution $\ha{1}$, which is stable for small $L$, becomes
unstable at $L=\sqrt{2}$. Then the $1$-skyrmion $\sk{1}$ 
(together with its reflections) is born by biffurcation from the 
$\ha{1}$ and exists as a stable solution for $L>\sqrt{2}$. 
In this unique case the
critical phenomenon may be astonishingly simply described
analytically and therefore completely understood. 
This gives rise to a
general picture of the process of the spontaneous breaking of
the reflection symmetry which may be intuitively presented as a
costing no energy inflation of the critical mode on the unstable
symmetric solution. The unstable solution and the critical mode
 have mutually opposite parity,
thus the newly born solution has spontaneously broken parity. 
Finnaly, we
also found two unique series expansions for the profile and
for the energy of the
$1$-skyrmion  (although the respective expansion
coefficients are not known in  general forms) which read
respectively
\begin{eqnarray*} && F(\psi)=\psi+x\sin{\psi}+x^2\frac{3}{20}\sin{2\psi}
-x^3\left(\frac{29369}{316800}\sin{\psi}-\frac{11}{480}\sin{3\psi}\right)\\
&&\qquad\qquad\qquad-x^4\left(\frac{4217}{887040}\sin{2\psi}
-\frac{109}{20160}\sin{4\psi}\right)
+\dots
\end{eqnarray*}
and
\begin{eqnarray*}
&& \frac{E[F]}{12\pi^2}=\frac{3\sqrt{2}}{4}\left(1+\frac{11}{180}x^2
-\frac{209}{10800}x^4+\frac{5209}{864000}x^6
 -\frac{2126641283}
{1149603840000}x^8+\olarge{(x^{10})}\right),
\end{eqnarray*}
where
$$
x=\pm\sqrt{\frac{60}{11}\left(\frac{L}{\sqrt{2}}-1\right)}.
$$
These series expansions are valid in some right neighbourhood of
the critical radius $L=\sqrt{2}$ and numerical assessment of their
behaviour makes them plausible to possesss nonzero radii
of convergence.

This critical behaviour which we observed in our model exist
neither in the Skyrme model on the flat space nor in the model of
harmonic maps between three-spheres when free parameters of these
models -- the characteristic size of a soliton in the first case,
or the radius of the base three-sphere in the second case -- are
being changed. It shows that by coupling together two field
theories we may produce another theory in which new phenomena may
appear that are absent in the decoupled case. Dimensional
parameters introduced by different models determine then some
dimensionless numbers whose values are usually crucial to the
existence of different types of solutions. This fact is known when
nonlinear fields are coupled to gravity \cite{bizchmaj}. The main
motivation of our work was to understand such model-independent
phenomena in possibly the simplest case which is a nonlinear
scalar field theory on a fixed space-time background.

%% file: jmp_app.tx
\section{Formal series expansion for the profile and energy of the
skyrmion $\sk{1}$ on $\s{3}$} \label{apdx:series} In order to find
a formal series expansion for the profile $F(\psi)$ of $\sk{1}$ in
the vicinity of $L=\sqrt{2}$ we assume that this solution can be
written using the following ansatz:
\begin{eqnarray}
&& F(\psi,x)=\psi \label{eq:profile}\\
&&+\left\{ a_1+x^2a_{3}\left(1+a_{3,3}
\frac{\sin{3\psi}}{\sin{\psi}}\right)+x^4a_{5}\left(1+a_{5,3}
\frac{\sin{3\psi}}{\sin{\psi}}+a_{5,5}\frac{\sin{5\psi}}{\sin{\psi}}\right)
+\dots\ \right\}\nonumber \\&&\qquad \cdot x\sin{\psi}\nonumber\\
&&+\left\{a_2+x^2a_4\left(1+a_{4,4}
\frac{\sin{4\psi}}{\sin{2\psi}}\right)+x^4a_6\left(1+a_{6,4}
\frac{\sin{4\psi}}{\sin{2\psi}}+a_{6,6}\frac{\sin{6\psi}}{\sin{2\psi}}
\right)+\dots\ \right\}\nonumber\\ &&
\qquad\cdot x^2\sin{2\psi}\nonumber
\end{eqnarray}where
$$x=\sqrt{\frac{60}{11}\left(\frac{L}{\sqrt{2}}-1\right)},\qquad L>\sqrt{2}, \quad
\mathrm{and} \quad a_1:=1.$$ This ansatz follows from the idea
that the solution $F(\psi,x)$ should be linearly decomposed in the
base of eigenfunctions of the Hessian (\ref{eq:Hessian})
evaluated at $\ha{1}$. This is implied by the appearance of
$\sk{1}$ at the critical radius $L=\sqrt{2}$ by the excitation of
the lowest energy mode (i.e. $\sin{\psi}$) on $\ha{1}$, which was
perfectly confirmed when we reproduced analytically the
numerical observation that
$$|a-1|=\sqrt{\frac{30\sqrt{2}}{11}\left(L-\sqrt{2}\right)}:=
x,$$
which implies that
$$ \frac{\partial
F(\psi,x)}{\partial{x}}\bigg|_{x=0}=\sin{\psi},$$ and the vague
idea that this behaviour should be somehow analytically continued
in the variable $x$. From the symmetry of equation (\ref{eq:main})
we impose on the series also the requirement that, since $L\propto
x^2$, the second solution $\pi-F(\pi-\psi,x)$ should be simply
$F(\psi,-x)$ (both skyrmions bifurcate from $\ha{1}$ at $x=0$ and
can be transformed one to another by the reflection with respect
to the point $(\pi/2,\pi/2)$).

The coefficients $a_k$ and $a_{k,l}$ (at least the one we had
found) are rational numbers and relations between them become more
and more complicated as $k$ increases. Due to nonlinearity of
equation (\ref{eq:main}) it seems there is no general recurrence
which would determine these coefficients. Anyway, the above ansatz
enables us to find them all successively step by step. In table
\ref{tab:coefficients} we give all the coefficients which are
required to determine the profile of $\sk{1}$ with the accuracy up
to $\olarge{(x^{14})}$. The question arises if the formal series
possesses nonzero radius of convergence. We gave in the main text
some arguments that this series is really a solution. In order to
enhance our argumentation we may try to estimate this radius. We
can treat the maximal absolute values of the functional
coefficients standing in braces at $x^{2k}$ in equation
(\ref{eq:profile}) to built a majorizing power series in the variable
$x$. Let denote the respective values by $\{c_1,c_3,\dots\}$ and
$\{c_2,c_4,\dots \}$ respectively for the first and the second
brace. Thus the partial sums of the above series are majorized by
$\sum c_n|x|^n$. In figure \ref{gr:profilerad}
\dblgr{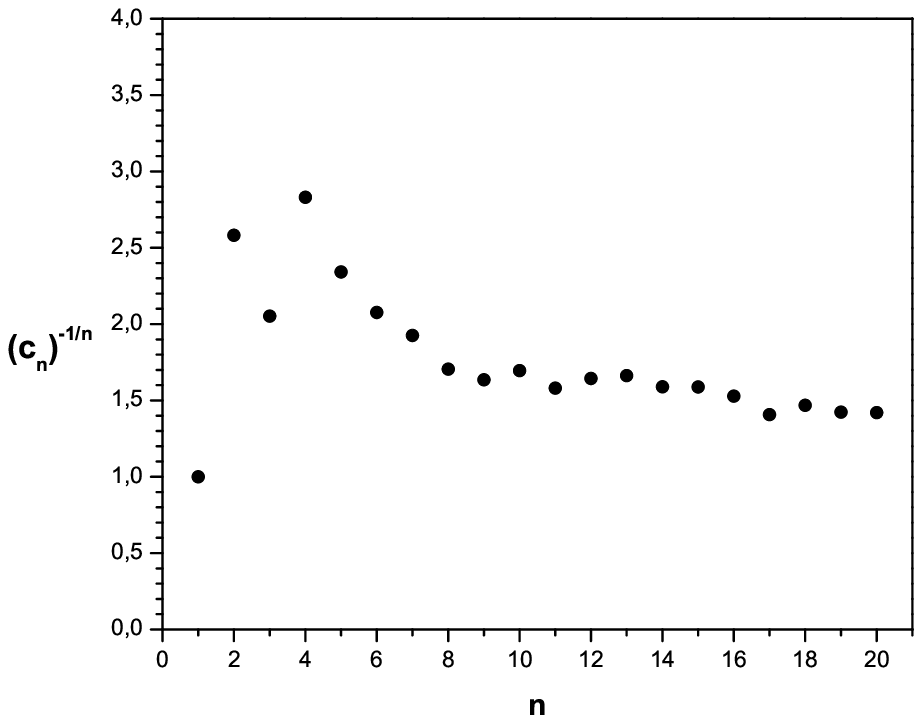} {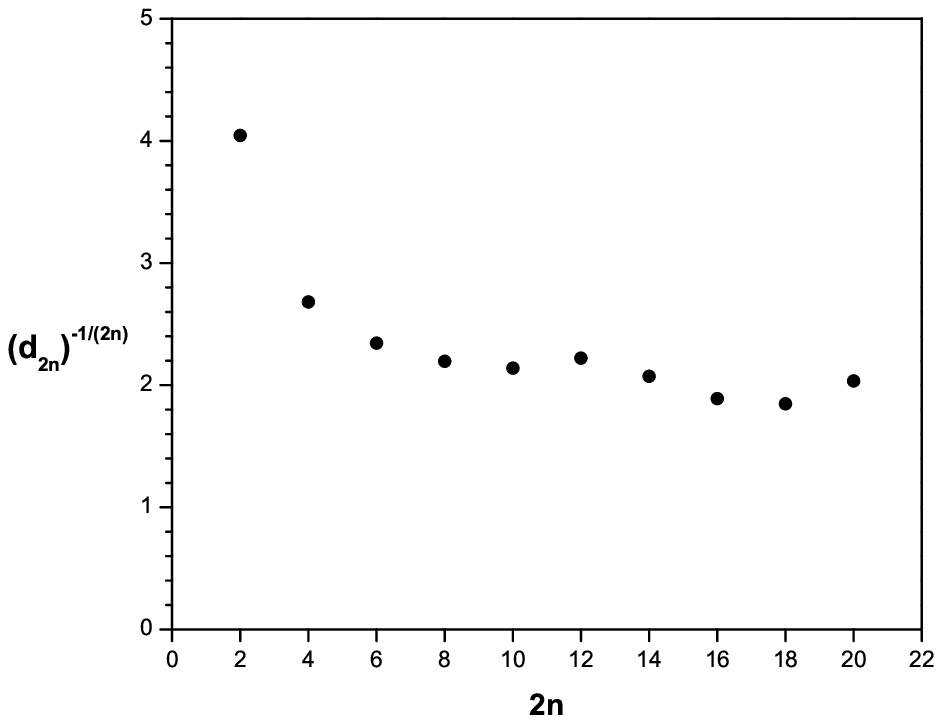} { \newline {\bf left:}
Initial elements of the sequence $\{1/\sqrt[n]{c_n}\}$, where
$\sum_nc_nx^n$ is a majorizing series for the profile of $\sk{1}$.
This strongly suggest that the Taylor-Fourier series expansion
(\ref{eq:profile}) has a nonzero radius of convergence and even
that, it is absolutely convergent. \newline {\bf right:} Initial
elements of the sequence $\{1/\sqrt[2n]{d_{2n}}\}$ where $d_{2n}$
are defined in (\ref{eq:energyexp}). This suggest that the Taylor
series expansion of energy of $\sk{1}$ may have a nonzero radius of
convergence.}{gr:profilerad}
are shown $20$ initial elements of the sequence
$(1/\sqrt[n]{c_n})$. Their behaviour strongly suggest the
hypothesis that the Taylor-Fourier series expansion
(\ref{eq:profile}) has a nonzero radius of convergence and even
that, it is absolutely convergent.
\begin{table}
\begin{small}
\begin{tabular}{ll}
\hline
    \spc
    $a_1=1$ & $a_{2}=+\frac{3}{20}$ \\
    \spc
    \hline
    \spc
    $a_{3}=-\frac{29369}{316800}$ & $a_{4}=-\frac{4217}{887040}$ \\
    \spc
    $a_{3,3}=-\frac{7260}{29369}$ & $a_{4,4}=-\frac{4796}{4217}$ \\
    \spc
    \hline
    \spc
    $a_{5}=+\frac{12335768813}{1405071360000} $ & $ a_{6}=+\frac{5010043861}{7376624640000}$ \\
    \spc
    $ a_{5,3}=-\frac{4404867060}{12335768813}$ & $a_{6,4}=-\frac{33475541308}{5010043861} $ \\
    \spc
    $a_{5,5}=+\frac{3226815900}{12335768813} $ & $ a_{6,6}=+\frac{13362918261}{10020087722}$ \\
    \spc
    \hline
    \spc
    $a_{7}=+\frac{7986485836639}{25751126016000000}$ & $a_{8}=-\frac{473077794653026103}{98150416809984000000}$\\
     \spc
    $a_{7,3}=+\frac{158986289932080}{7986485836639}$ & $a_{8,4}=-\frac{361662885996135944}{473077794653026103}$\\
     \spc
    $a_{7,5}=-\frac{57715917199500}{7986485836639}$ & $a_{8,6}=+\frac{51442637682491484}{473077794653026103}$\\
     \spc
    $a_{7,7}=+\frac{6755487649680}{7986485836639}$ & $a_{8,8}=-\frac{5773312055063520}{473077794653026103}$\\
    \spc
    \hline
    \spc
    $a_{9}=-\frac{127577656118454895711}{167510044689039360000000}$ & $a_{10}=+\frac{4211381544228526601549071}{1061084526049375027200000000}$\\
    \spc
    $a_{9,3}=+\frac{790229840916734772664}{127577656118454895711}$ & $a_{10,4}=-\frac{822641146563016022074558}{4211381544228526601549071}$\\
    \spc
    $a_{9,5}=-\frac{309555559736002696600}{382732968355364687133}$ & $a_{10,6}=-\frac{1447811040980803898316621}{33691052353828212812392568}$\\
    \spc
    $a_{9,7}=+\frac{20023176097891467424}{382732968355364687133}$ & $a_{10,8}=-\frac{13895432790539792801760}{4211381544228526601549071}$\\
    \spc
    $a_{9,9}=-\frac{2544822818150368800}{127577656118454895711}$ & $a_{10,10}=+\frac{55236971873923883695575}{33691052353828212812392568}$\\
    \spc
    \hline
    \spc
    $a_{11}=+\frac{982955548821103176200169976183}{2390411220284032061276160000000000}$ & $a_{12}=-\frac{107845172239555936226487785603477}{73415504602973334681944064000000000}$\\
    \spc
    $a_{11,3}=+\frac{3024748425995528913667880277780}{982955548821103176200169976183}$ & $a_{12,4}=+\frac{60201002753190711583565881906936}{107845172239555936226487785603477}$\\
    \spc
    $a_{11,5}=+\frac{1248755928443601042244324285400}{982955548821103176200169976183}$ & $a_{12,6}=-\frac{8388700062249632787593687120376}{107845172239555936226487785603477}$\\
    \spc
    $a_{11,7}=-\frac{58074313732006509362657835680}{982955548821103176200169976183}$ & $a_{12,8}=-\frac{5165826035166448141662612158400}{107845172239555936226487785603477}$\\
    \spc
    $a_{11,9}=-\frac{56310543306011901706414401600}{982955548821103176200169976183}$ & $a_{12,10}=+\frac{975465355644891981999033888060}{107845172239555936226487785603477}$\\
    \spc
    $a_{11,11}=+\frac{6823246116254206530384345600}{982955548821103176200169976183}$ & $a_{12,12}=-\frac{68171833635114106739516947272}{107845172239555936226487785603477}$\\
    \spc
    \hline
    \spc
    $a_{13}=-\frac{43153404274707851225466988047569832167}{413476121923945820928708968448000000000000}$ & $a_{14}=-\frac{1766865169954398109033856353883942537617}{6132984423523186432660243218432000000000000}$\\
     \spc
    $a_{13,3}=-\frac{313572180766907863870688164387239391620}{43153404274707851225466988047569832167}$ & $a_{14,4}=-\frac{4349592583735535844513687814933733037976}{1766865169954398109033856353883942537617}$\\
     \spc
    $a_{13,5}=+\frac{137371166634494138333962660336190541100}{43153404274707851225466988047569832167}$ & $a_{14,6}=-\frac{1508853846688227478046216373401843325339}{1766865169954398109033856353883942537617}$\\
     \spc
    $a_{13,7}=+\frac{65940526790056736560082171865362030240}{43153404274707851225466988047569832167}$ & $a_{14,8}=+\frac{585277715345242542258046459394514314520}{1766865169954398109033856353883942537617}$\\
     \spc
    $a_{13,9}=-\frac{16887408629502465138357459627899436000}{43153404274707851225466988047569832167}$ & $a_{14,10}=-\frac{78074347310444368596245949472056202125}{3533730339908796218067712707767885075234}$\\
     \spc
    $a_{13,11}=+\frac{1437225091006557895313712800717160000}{43153404274707851225466988047569832167}$ & $a_{14,12}=+\frac{1726765295922525300215513807655077490}{1766865169954398109033856353883942537617}$\\
     \spc
    $a_{13,13}=-\frac{92252259291662946944254093579272000}{43153404274707851225466988047569832167}$ & $a_{14,14}=-\frac{38714472844152751765544208844685625}{199083399431481477074237335648894933816}$\\
    \spc
    \hline
\end{tabular}
\end{small}
\caption{Initial coefficients of the Taylor-Fourier series
expansion (\ref{eq:profile}) of the solution $\sk{1}$ in the
vicinity of the critical radius
$L=\sqrt{2}$.}\label{tab:coefficients}
\end{table}

The Taylor series expansion of the energy of $\sk{1}$ 
around $x=0$ must contain only even
powers of $x$ since $F(\psi,x)$ and $F(\psi,-x)$ must have equal
energies. Therefore the  energy of $\sk{1}$ is a smooth function of 
 $L-\sqrt{2}$,
unlike the $F'(0,x(L))-1$, which has a branch singularity at
$L=\sqrt{2}$. This series reads
\begin{eqnarray}
\frac{E(x)}{12\pi^2}=\frac{3}{4}\sqrt{2}\left(1+d_2x^2+d_4x^4+\dots
\right) \label{eq:energyexp}
\end{eqnarray}
 and the respective coefficients are given in table
\ref{tab:energyexp}.
\begin{table}
\begin{small}
\begin{tabular}{l} \hline
\spc $d_{2\ }=+\frac{11}{180}$\\
\spc $d_{4\ }=-\frac{209}{10800}$\\
\spc $d_{6\ }=+\frac{5209}{864000}$\\
\spc $d_{8\ }=-\frac{2126641283}{1149603840000}$\\
\spc $d_{10}=+\frac{8434513426021}{16995743170560000}$\\
\spc $d_{12}=-\frac{464402526814953371}{6730314295541760000000}$\\
\spc $d_{14}=-\frac{23242947675020983465203089}{626856089235323092992000000000}$\\
\spc $d_{16}=+\frac{14367315331877365555523427348253}{379122562769523406641561600000000000}$\\
\spc $d_{18}=-\frac{31939951417910304147919969132085999}{2017781268474468255763984809984000000000}$\\
\spc $d_{20}=-\frac{74250563114418365106618380718783299763796141}{109069148686118907097066434918875136000000000000000}$\\
\spc \hline
\end{tabular}
\end{small}
\caption{Initial coefficients of the Taylor series expansion of
energy of $\sk{1}$ at the critical radius $L=\sqrt{2}$. In order
to calculate $d_{2n}$ we need the profile of $\sk{1}$ to be
determined with the accuracy up to $x^{2n}$.}
\label{tab:energyexp}
\end{table}
For $x=0$ we get the energy of $\ha{1}$ at $L=\sqrt{2}$. Here we
also construct the sequence $\{1/\sqrt[2n]{d_{2n}}\}$ to make it
plausible that the series (\ref{eq:energyexp}) and, indirectly, the
series (\ref{eq:profile}), have nonzero radius of convergence. In
figure \ref{gr:profilerad}  are shown initial elements of the
sequence $(1/\sqrt[2n]{d_{2n}})$. Their behaviour suggest the
hypothesis that the Taylor series expansion (\ref{eq:energyexp})
may have a nonzero radius of convergence.

\section{On the weight function originating from a kinetic term}
\label{apdx:eigen} 

In this appendix we derive the formula for the
weight function, which naturally follows from the calculus of
small oscillations about equilibrium. We obtain this by the
comparison of two conceptually different approaches to linear
stability analysis. Suppose, for example, that we have a
differential equation for a function of one space variable and
that the equation is a condition for extrema of some functional.
If the function is a solution and, moreover, a local minimum of
the functional then it is also a stable equilibrium of some $1+1$
dimensional field theory. Stability is required for physical
reasons. In order to check that the function is a minimum it
suffices to prove positive definiteness of the second variation of
the functional. This in turn requires the existence of a scalar
product (which should be defined by additional argumentation) and
defines a Sturm-Liouville problem. On the other hand, in linear
approximation, the calculus of small oscillations about the
equilibrium produces another Sturm-Liouville problem with a
definite weight function determined by the kinetic term of the
field theory. It turns out that both the problems comprise
identical self-adjoint differential operators (the  function
spaces of admissible perturbations are the same by construction).
Thus, to specify the weight function in the first case, it is
natural the requirement that both methods should give the same
results. Of course, to check positive definiteness of the second
variation of the functional we can assume an arbitrary weight
function. If the function is not a minimum we are not guaranteed
that the different field theories have qualitatively the same
dynamics in the vicinity of the common equilibrium. It would be
so, under the condition that the number of instable modes were
independent of the weight functions, which is not obvious. In what
follows we determine the weight function.

We can supplement functional
(\ref{eq:energy}) with an arbitrary kinetic term not depending
explicitly on time (to have the total energy
$\mathcal{E}[F(\tau)]$ conserved)$$ \mathcal{E}[F(\tau)]=\int
\limits_{0}^\pi w(\psi)\left(
K(\psi,F(\psi,\tau),\dot{F}(\psi,\tau))+
V(\psi,F(\psi,\tau),F'(\psi,\tau))\right)\ud{\psi},
$$ and observe the resulting time-evolution of some initial data in time $\tau$.
Conventionally we measure duration using natural units of length
$\e^{-1}\f^{-1}$ then $\tau $ is dimensionless.  The quantities
$K$ and $V$ are interpreted as densities of kinetic and potential
energy respectively. We assume that $K=\tilde{K}\dot{F}^2$ where
$\tilde{K}$ is positive function of $\psi,F,F',\dot{F}$. A
perturbation  can be imagined as a time dependent excess from the
equilibrium. An equilibrium solution $F$ is said to be dynamically
stable if it remains within some finite bounds $\norm{\delta
F(\tau)}{}<\varepsilon(\Delta)$ for every $\tau$ if only the
initial perturbation is sufficiently small such that $\delta
E<\Delta$. Thus, to check stability an initial solution is
required to be not arbitrary but such, as its total energy was
slightly above the energy of the equilibrium solution and in its
pointwise vicinity. It is clear that different kinetic terms give
rise to different time evolution and the equilibria are common and
the same as solutions of equation (\ref{eq:main}), since the
condition that $\delta{\mathcal{U}[F]}/\delta{F(\psi)}=0$ for
static solutions is the straightforward consequence of equation of
motion
\begin{equation}\frac{\partial{K}}{\partial{F}}-\frac{\ud}{\ud{t}}
\frac{\partial{K}}{\partial{\dot{F}}}-
\frac{\delta{\mathcal{U}[F]}}{\delta{F}}=0\label{eq:kinetic}\end{equation}
and of the requirement that the solutions were in equilibrium. If
we had decided to assume the kinetic energy density as in the
original Skyrme model on $\s{3}$, i.e.
\begin{equation}K(\psi,F,\dot{F})=\left(L+\frac{2}{L}\frac{\sin^2{F}}
{\sin^2{\psi}}
\right)L^2\dot{F}^2,\label{eq:skyrmkinetic}\end{equation} then the
evolution of the system would have been governed by the equation
\begin{eqnarray*}\left(L\sin
^2{\psi}+\frac{2}{L}\sin^2{F}\right)\left(F''-L^2\ddot{F}\right)
+\frac{1}{L}\sin{2F}\left(F'^2-L^2\dot{F}^2\right)+L\sin{2\psi}F'
\\-\left(L+\frac{1}{L}\frac{\sin ^2{F}}{\sin
^2{\psi}}\right)\sin{2F}=0,
\end{eqnarray*}
while assuming $K(\psi,F,\dot{F})=L^3\dot{F}^2$ we would have been
 led to
 the equation
\begin{eqnarray*}\left(L\sin
^2{\psi}+\frac{2}{L}\sin^2{F}\right)F''- \sin ^2{\psi}L^3\ddot{F}
+\left(L+\frac{1}{L}\frac{\sin{2F}}{\sin{2\psi}}F'\right)\sin{2\psi}
F'\\-\left(L+\frac{1}{L} \frac{\sin ^2{F}}{\sin
    ^2{\psi}}\right)\sin{2F}=0.\end{eqnarray*}
As it will turn out the last choice for $K$ gives, up to a
constant factor $L^3$, the natural on $\s{3}$ weight function
$4\pi\sin^2{\psi}$.
We can always look for solutions in the form of
$F(\psi,\tau)=F(\psi)+\varepsilon \eta(\psi,\tau)$ where
$\varepsilon$ measures the scale of perturbation, i.e. its energy,
which is proportional to $\varepsilon^2$.  Next, for small
$\varepsilon$ by analogy with the calculus of small oscillations,
solutions (if exist) for which $\eta$ is bounded for all $\tau$,
may be arbitrarily well approximated by the following linear
equation, which is derived by equating with zero the term of the
order $\olarge{(\varepsilon)}$ in the Taylor series expansion of
equation (\ref{eq:kinetic}) in the variable $\varepsilon$, i.e.
\begin{equation} -\left(w\frac{\partial^2U}{\partial F'^2}
\frac{\partial{\eta}}{\partial{\psi}}\right)'+\left(w\frac{\partial^2U}
{\partial F^2}- \left(w\frac{\partial^2U}{\partial
FF'}\right)'\right)\eta=-
w\frac{\partial^2K}{\partial\dot{F}^2}\frac{\partial^2\eta}{\partial\eta^2}.
\label{eq:eig2}\end{equation}
 Other terms (not shown) which enter above equation vanish since we perturb
  a static solution for which
  $\dot{F}=0$. If the
equilibrium was not stable the approximation would still have been
useful to predict the existence of instabilities. It follows, that
for the class of kinetic terms assumed it is always possible to
find $\eta$ by the method of separation of variables
$\eta(\psi,\tau)=\xi(\psi)\alpha(\tau)$, where $\alpha$ is of the
form $\alpha(\tau)=e^{i\omega \tau}$ and $\omega^2<0$ for unstable
modes. Thus the dynamics impose the natural choice that weight
function $g(\psi)$ should be defined according to the formula
\begin{equation}g(\psi)=w(\psi)\frac{1}{2}
\frac{\partial^2K}{\partial\dot{F}\partial\dot{F}}\bigg|_{F_{equil}},
\label{eq:kineticnorm}\end{equation} since equation
(\ref{eq:kineticnorm}) reduces to equation (\ref{eq:eig})  after
the substitution
$\eta(\psi,\tau)=\xi(\psi)\sin{(\omega\tau+\phi)}$ with
$\lambda=\omega^2$ or, if the mode is unstable,
$\eta(\psi,\tau)=\xi(\psi)\exp{(\pm\omega\tau)}$ with
$\lambda=-\omega^2$. Put differently, multiplying (\ref{eq:eig2})
by $\eta/2$ and integrating, yields
$$\frac{1}{2}\hess=\lambda\int\limits_0^{\pi}w\frac{1}{2}
\frac{\partial^2K}{\partial\dot{F}\partial\dot{F}}\xi^2$$ since
$\ddot{\alpha} +\lambda\alpha=0$ which reproduces (\ref{eq:eig})
for the special choice of $g$. In this sense, the spectrum of
(\ref{eq:eigen}) with $g=w$ may be interpreted as $\omega^2$ for a
spherically symmetric scalar field on $\s{3}$ in an equilibrium if
$K=\dot{F}^2$. Thus, the Hilbert space spanned on vanishing on
boundaries and mutually orthogonal (with respect to the special
$g$) eigenfunctions $\xi_i$  of  the self-adjoint equation
(\ref{eq:eig2}) is constructed and each perturbation $\eta$ is a
superposition
$\eta(\psi,\tau)=\sum_k\xi_k(\psi)(a_k\alpha_k(\tau)+b_k\beta_k(\tau))$
where $\alpha_k$ denotes $\sin{\omega_k\tau}$ or
$\sinh{\omega_k\tau}$ and $\beta_k$ denotes $\cos{\omega_k\tau}$
or $\cosh{\omega_k\tau}$, respectively, if $\lambda_k$ is positive
or negative. Only a finite number of modes can be unstable. The
total energy of the perturbation is finite and quadratic in
amplitudes
   (proportional to $\varepsilon^2$) and, of course, only if all $\lambda_k$
   are positive  the function  $F(\psi)+\varepsilon\eta(\psi,\tau)$
well approximates the particular solution of equation
(\ref{eq:kinetic}) for all $\tau$.

It is thus seen that the linear dynamical stability is closely
related to the (abstract) notion of linear energetical stability
defined in \ref{dfn:stability} and, if the weight function $g$ is
appropriately chosen, all eigenvalues of the Hessian $\hess$ may
be directly interpreted as squared eigenfrequencies $\omega_k^2$
of a system whose dynamics is related to some kinetic term.